\documentclass[11pt]{article}
\usepackage{xcolor}
\usepackage[utf8]{inputenc}
\usepackage{physics}
\usepackage{booktabs}
\usepackage{amsmath,bm,amssymb}
\usepackage[table]{xcolor}
\usepackage{siunitx}
\usepackage{graphicx}

\usepackage{float}
\usepackage{placeins}
\usepackage[letterpaper,margin=1in]{geometry}

\renewenvironment{abstract}
	{\quotation}
	{\endquotation}

\date{}

\makeatletter
\renewcommand{\fnum@figure}{\textbf{Figure \thefigure}}
\renewcommand{\fnum@table}{\textbf{Table \thetable}}
\makeatother

\usepackage{url}

\def\scititle{
	Chiral-phonon generation of orbital currents in light transition metals
}
\title{\bfseries \boldmath \scititle}

\author{
Marc Rovirola$^{1,2,*}$,
J\'ulia \`Odena$^{1}$,
Anna Castellv\'{\i}$^{1}$,
Quim Badosa$^{1}$,
Blai Casals$^{3,2}$\\
Adri\'an Gud\'{\i}n$^{4,5}$,
Haripriya Madathil$^{4}$,
Fernando Ajejas$^{5}$,
Paolo Perna$^{5}$\\
Alberto Hern\'andez-M\'{\i}nguez$^{6}$,
Joan Manel Hern\`andez$^{1,2}$,
Sa\"ul V\'elez$^{4,7,\ddagger}$,
Ferran Maci\`a$^{1,2,\dagger}$\\[6pt]
\small $^{1}$Dept.\ of Condensed Matter Physics, University of Barcelona, 08028 Barcelona, Spain\\
\small $^{2}$Institute of Nanoscience and Nanotechnology (IN2UB), University of Barcelona, 08028 Barcelona, Spain\\
\small $^{3}$Dept.\ of Applied Physics, University of Barcelona, 08028 Barcelona, Spain\\
\small $^{4}$IFIMAC and Departamento de F\'isica de la Materia Condensada, Universidad Aut\'onoma de Madrid, \\ 28049 Madrid, Spain\\
\small $^{5}$IMDEA Nanociencia, 28059 Madrid, Spain\\
\small $^{6}$Paul-Drude-Institut f\"ur Festk\"orperelektronik, 10117 Berlin, Germany\\
\small $^{7}$Instituto Nicol\'as Cabrera, Universidad Aut\'onoma de Madrid, 28049 Madrid, Spain\\[6pt]
\small $^\ast$marc.rovirola@ub.edu \,
\small $^\ddagger$saul.velez@uam.es \,
\small $^\dagger$ferran.macia@ub.edu
}

\begin{document} 

\maketitle

\begin{abstract} \bfseries \boldmath
Orbital angular momentum offers a new channel for information transport in a vast set of materials. Its coherent generation and detection remain, however, largely unexplored. Here, we demonstrate that chiral surface acoustic waves (SAWs) generate sizable orbital currents in light-metal/ferromagnet bilayers through both the acoustic orbital Hall effect and acoustic orbital pumping. Using symmetry analysis of SAW-driven voltages, we disentangle vorticity-sensitive orbital currents arising from lattice rotation in the non-magnetic layer from angular-momentum pumping from the ferromagnet. Strong signals are observed only in nickel$|$chromium and nickel$|$titanium, while nickel$|$aluminum and all cobalt-based bilayers show negligible responses, revealing the critical roles of orbital Hall conductivity, phonon–orbital coupling, and interfacial orbital transparency. Comparison with spin-torque ferromagnetic resonance and second-harmonic measurements---where electrically driven orbital angular momentum are weaker---demonstrates that phonon excitation generates orbital currents more efficiently. These results establish chiral SAWs as an effective route for orbitronic functionality and open pathways toward phonon-controlled orbital magnetism.
\end{abstract}

\section*{Introduction}
\noindent
While electron spin has dominated magnetic device research---with birth and maturity of spintronics---its orbital counterpart remains largely unexplored. The orbital degree of freedom is ubiquitous across all materials and might have stronger couplings to electric fields and lattice distortions, offering a rich playground for novel functionalities. Exploiting orbital angular momentum in solid-state systems could therefore expand the toolkit of magneto-electronic devices, complementing spin–based approaches with new mechanisms for control, detection, and energy conversion.

Theoretically, orbital currents arise from the orbital Hall effect (OHE), in which a charge current generates a transverse flow of orbital angular momentum even in light metals, and may exert “orbital torques” \cite{OHE_Texture_prl2018,Salemi_PRM2022,Choi2023}. Experimentally, sizable OHE signals have been observed in several metals \cite{Gambardella_prb_OHE_2022}, and the reciprocal effects have enabled the detection of orbital currents generated by magnetization-driven orbital pumping \cite{Hayashi2024_IOHE_orbital_pumping}. 

Phonons---the quanta of lattice vibrations---offer a fresh pathway into this untapped degree of freedom: by using surface acoustic waves (SAWs) to launch coherent phonons, one can drive magnetization dynamics through the magneto‐elastic effect \cite{weiler2011elastically,casals2020generation,SAW_Co_Ni_Rovirola} and also directly pump spin or orbital angular momentum from a ferromagnet into an adjacent non‐magnetic layer. Experimental studies reported this effect in bilayered systems of ferromagnets in contact with heavy elements such as platinum \cite{Weiler_CoPt,Rovillain_CoPtSAW}, tantalum \cite{Ni_Ta_SAW_spin_rotation}, tungsten \cite{Spin_rotation_W_Pt_SAW,Acoustic_spin_Hall_SO_saw} or bismuth \cite{edelstein_saw_Bi_prb,Puebla_2020_SAW_BiO}. More recently, it has been shown that the rotational component of the SAW itself can generate angular-momentum currents even in non-magnetic metals: chiral phonons can create transverse spin currents via the acoustic spin Hall effect, ASHE\cite{Acoustic_spin_Hall_SO_saw}, and, more interestingly, orbital currents via the acoustic orbital Hall effect, AOHE \cite{acousticOHE2025}. Thus, lattice rotation alone, independent of magnetic dynamics, may act as an efficient source of orbital angular momentum in solids.

Our experiment quantifies phonon-driven orbital currents in ferromagnet/light-metal bilayers. SAWs are used both to excite magnetoacoustic dynamics that enable orbital pumping from the ferromagnet and to generate chiral lattice motion in the light metal that produces an orbital current, allowing us to detect both mechanisms simultaneously. We compare caps with strong versus negligible orbital‐to‐charge conversion---chromium (Cr)/titanium (Ti) versus aluminum (Al)---and contrast two ferromagnets: nickel (Ni), which efficiently emits orbital currents, and cobalt (Co), which does not.
Our measurements reveal large, vorticity-sensitive orbital currents in Ni|Cr and Ni|Ti, whereas Ni|Al and Co-based devices produce minimal signals. Comparisons with electrically generated orbital currents in the same samples indicate that SAW-driven orbital generation is much more efficient.

\section*{Results}
\subsection*{Experimental setup}
\begin{figure}[ht]
    \centering
    \includegraphics[width=1\textwidth]{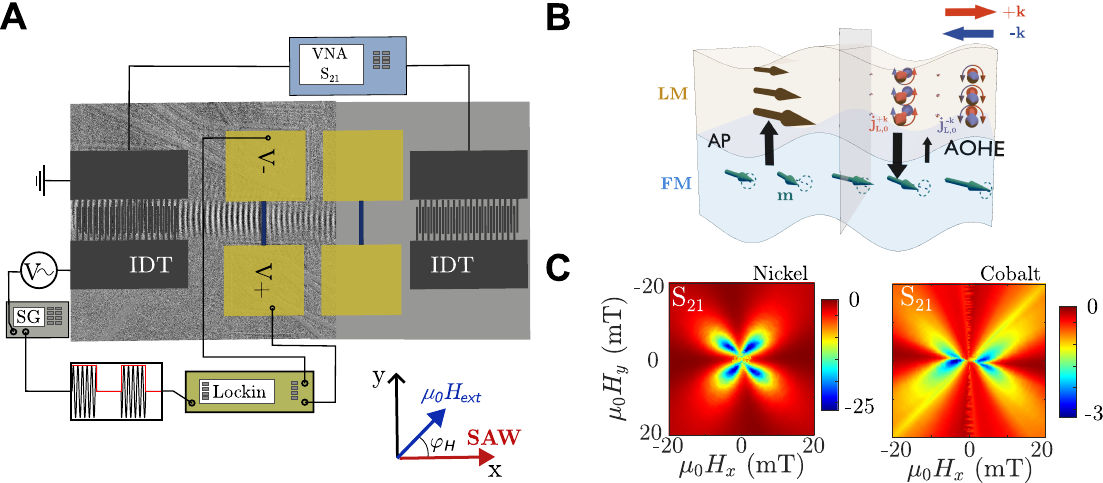}
    \caption{(\textbf{A}) Schematic diagram of the experimental setup. The device consists of a LiNbO$_3$ substrate with two opposing IDTs that can generate SAWs propagating with opposite wavevectors. A FM/LM bilayer (FM = Ni, Co, and LM = Cr, Al) patterned in a bar geometry is grown perpendicular to the SAW propagation direction and in the acoustic path. Gold contacts are grown on each end of the bar to detect acoustically induced voltages. An in-plane magnetic field is applied with an electromagnet, and the sample can be rotated within the field plane. Transmission measurements ($S{_{21}}$) are performed with a Vector Network Analyzer (VNA) at the resonant IDT frequencies as a function of magnetic field, while voltage measurements are carried out by exciting the IDT with low-frequency amplitude modulated microwave pulses and detected the synchronized response with a lock-in amplifier.
    (\textbf{B}) Schematics of acoustic pumping (AP) (left-side), where a spin or orbital current is pumped into an adjacent non-magnetic layer and transformed into a charge current through ISHE or IOHE.  Schematics of acoustic orbital hall effect (AOHE) (right-side) rising from the SAW lattice rotations coupled to the orbital degree of freedom in the LM, generating an orbital current polarized in the $\mathbf{y}$ direction. This current can interact with the magnetization of the FM, with a modulation that depends on their relative orientation. 
    (\textbf{C}) Acoustic-FMR transmission measurements ($S_{21}$) at 1.3 GHz    of Ni (left) and Co (right) displaying the expected 4-fold symmetry and resonance peaks.}
    \label{setup}
\end{figure}

The samples under study consist of piezoelectric 127.86$^\circ$ $Y$-cut LiNbO$_3$ substrates and bilayers of 10-nm-thick ferromagnet interfaced with a 10-nm-thick non-magnetic metal and capped with a 3-nm Al, all deposited by e-beam evaporation and patterned on the acoustic path with different shapes (See, Fig.\ \ref{setup}). Unidirectional IDTs were patterned at opposite ends of the bilayer structures, allowing for the generation of SAWs that propagate forward ($+k$) and backward ($-k$). The spatial periodicity of the finger-like electrodes in the IDTs was designed to excite SAWs with different harmonics, ranging from hundreds of MHz up to 2 GHz. The heterostructures allow for both measurements of transmitted SAW power ($S_{21}$ coefficient) and measurements of induced voltages generated across the samples. We designed the IDTs to generate Rayleigh SAWs, characterized by in-plane and out-of-plane longitudinal displacement components having a $\pi/2$ shift and causing a physical rotation of the lattice. By adjusting the relative amplitudes of counter-propagating SAWs from opposite IDTs, one can continuously tune between propagating and standing wave regimes, effectively switching from forward ($+k$) to backward ($-k$) propagation through $k = 0$ \cite{StandingSAW2019}. More details on sample growth, patterning, and characterization can be found in the Supplementary Note 1. Figure\ \ref{setup}a shows an optical microscope image of one of the measured samples showing the two-faced IDTs with a 10 $\mu$m wide FM/NM stripe with gold contacts. SAWs of 200 MHz are imaged using stroboscopic Kerr \cite{McCord_AEM_2022} and clearly show the acoustic path.

\begin{figure}[ht]
    \centering
    \includegraphics[width=0.6\textwidth]{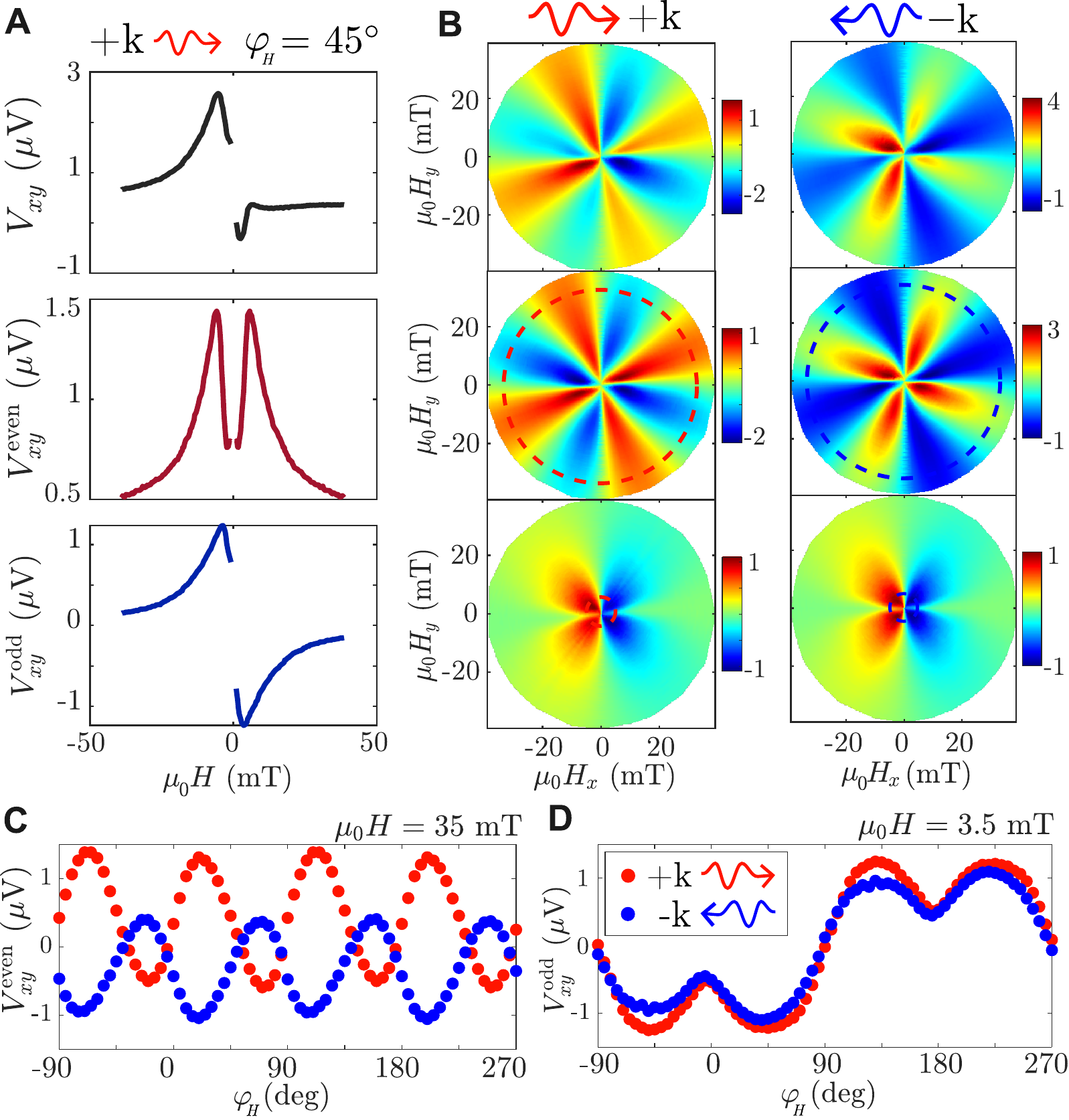}
    \caption{
    of transverse voltage contributions in Ni/Cr.
    (\textbf{A}) Representative voltage signals measured as a function of the magnetic field ($\mu_0 H$) at an angle $\varphi_H=45^\circ$, showing the total (top), even (middle), and odd (bottom) contributions. The magnetic field sweeps from a large magnetic field to zero.
    (\textbf{B}) Two-dimensional maps of the total (top), even (middle), and odd (bottom) voltage components as a function of in-plane ($\mu_0H_x$ and $\mu_0H_y$) magnetic fields. The left column shows the results for a SAW traveling in the positive x-direction, and the right column shows the results for a SAW traveling in the negative x-direction.
    (\textbf{C}) Angular dependence of the even voltage component at $\mu_0H = 35$ mT for both $+k$ (blue) and $-k$ (red) SAWs. \textbf{(d)} Angular dependence of the odd voltage components at $\mu_0H = 3.5$ mT for both $+k$ (blue) and $-k$ (red) SAWs.}
    \label{fig:orbitals}
\end{figure}
\vspace{0,0cm}
A GHz rf signal is applied to one of the two IDTs, generating SAWs in the piezoelectric substrate. The resulting oscillating strain may excite magnetoacoustic waves \cite{casalsGenerationImagingMagnetoacoustic2020,rovirola2023_physrevapp} in magnetostrictive materials \cite{yang2021acoustic,puebla2022perspectives}. While sweeping an external in-plane magnetic field $\mathbf{H}$ (with angle $\varphi_H$ relative to SAW propagation direction), the system may reach an acoustic ferromagnetic resonance condition at which part of the SAW energy is transferred to the magnetic system. This coupling leads to a measurable attenuation of the SAW amplitude \cite{weiler2011elastically} (see Fig.~\ref{setup}c). For a Rayleigh wave that travels along the $x$-direction, the magnetoelastic interaction can be phenomenologically described as
\begin{equation}
\begin{split}
    F_{me} = B_1(e_{xx}m_{x}^2+e_{zz}m_{z}^2) + 2B_2e_{xz}m_xm_{z},
\end{split}
\end{equation}

\noindent where $B_i$ are the magnetoelastic constants that relate the effect of strain on magnetization, $e_{ij}$ is the strain tensor coefficients, and $m_i$ are the normalized magnetization components. The corresponding magnetoelastic effective field is obtained as $\mu_0\mathbf{h}_{me} = -\nabla_{\mathbf{m}} F_{me}$ which contains terms proportional to $\sin(2\varphi_H)$ (see Supplementary Note 3). This angular dependence implies that maximum torque, and by extension, maximum energy absorption, occurs at an angle $\varphi_H$ of about 45 degrees with the SAW propagation direction \cite{weiler2011elastically,gowthamTravelingSurfaceSpinwave2015b,Labanowski2016}. Figure\ \ref{setup}c shows SAW attenuation as a function of the magnitude and angle of the applied magnetic field for a Ni (left panel) and Co (right panel) samples (tiny non-reciprocal effects are also observed for the measured frequencies \cite{Alberto_2020,Xu_magnetorotation_2020,GiantNonReciprocalShah2020}). 

\subsection*{Acoustic voltage signal decomposition}
We recorded the transverse dc voltage induced in the stripe heterostructures in the presence of a 1.3 GHz SAW. On the one hand, the precessing magnetization may pump angular‐momentum currents (spin or orbital) into the adjacent cap layer \cite{Hayashi2024_IOHE_orbital_pumping,Weiler_CoPt,Rovillain_CoPtSAW,Ni_Ta_SAW_spin_rotation,Spin_rotation_W_Pt_SAW,Acoustic_spin_Hall_SO_saw,edelstein_saw_Bi_prb,Puebla_2020_SAW_BiO} which can be converted to a dc signal via inverse Hall effects \cite{prb_spinHall_3dmetals_2014,PRL2022_MR_rashba_orbitals}. The signal arising from angular-momentum pumping depends on the magnetization direction and thus must reverse with the magnetic-field direction, remaining independent of SAW propagation direction. On the other hand, orbital or spin currents generated in the non-magnetic layer due to phonon-orbital effects may interact with the magnetization in the magnetic layer and eventually convert to dc signals through the inverse orbital or spin Hall effects (IOHE/ISHE). This signal must be independent of magnetic-field sign and reverse with SAW propagation direction \cite{Acoustic_spin_Hall_SO_saw}. 
Additional rectified voltages might also arise from the stripe’s resistance oscillations combined with the leaky electromagnetic waves that accompany the SAW \cite{SAW_electrical_adv_mat2023}; those signals must also be independent of the magnetic-field's sign. 


A representative voltage curve as a function of the applied magnetic field at $\varphi_H=45$ degrees shows strong asymmetric resonance peaks (Fig.\ \ref{fig:orbitals}a (top)). The curve can be decomposed into even (field-symmetric) and odd (field-antisymmetric) components (see Materials and Methods for details), which are shown in the middle and lower panels of Fig.\ \ref{fig:orbitals}a. We repeated the procedure for each field orientation and combined it into a color plot in Fig.\ \ref{fig:orbitals}b---both SAW propagation directions ($\pm k$) are included. Finally, panels (c) and (d) show the angular dependence at fixed magnetic field for even (at 35 mT) and odd (at the resonance field) components in both SAW propagation directions. The even component is shown at large field values to avoid resonance, where a phase shift occurs between the magnetization dynamics and the SAW-induced orbital currents in the non-magnetic layer (See, Supplementary note 6). The SAW propagation ($\pm k$) induces an inversion of the even component and no variation in the odd one, as previously shown in acoustic spin pumping experiments \cite{Hayashi2024_IOHE_orbital_pumping,Weiler_CoPt,Rovillain_CoPtSAW,Ni_Ta_SAW_spin_rotation,Spin_rotation_W_Pt_SAW,Acoustic_spin_Hall_SO_saw,edelstein_saw_Bi_prb,Puebla_2020_SAW_BiO}.

We associate the odd and even contributions to orbital pumping from the ferromagnet and to the acoustic orbital currents generated in the non-magnetic light metal overlayer, respectively. The oscillating part of the even contribution may correspond to the recently studied acoustic orbital Hall effect \cite{acousticOHE2025}, the orbital counterpart of the acoustic spin Hall effect \cite{Acoustic_spin_Hall_SO_saw}.

\subsection*{Acoustic orbital Hall effect (AOHE)}
Focusing first on the even contribution: the AOHE \cite{wuAcousticOrbitalHall2025,acousticOHE2025} links the lattice rotational motion caused by the SAWs and the angular momentum of electrons and can generate orbital currents even in non-magnetic materials or in materials with low spin-orbit coupling. The generated orbital current is set by the local lattice vorticity and the angular velocity induced by the SAWs, and therefore oscillates in both space and time (see Fig.,\ref{setup}b and Supplementary Material). The resulting orbital flow, perpendicular to the film along the $z$-direction, is initially polarized along $y$ and can be written as:


\begin{equation}
    j_{L,0}^{y} = -A k \omega u_0\sin\!\big(kx - \omega t\big),
    \label{eq:jl0}
\end{equation}
\noindent where $A$ determines the efficiency of generating this orbital current, which encompasses the phonon-orbital coupling strength ($\chi_{po}$), $u_{0}$ is the amplitude of the SAW, $\omega$ is the angular frequency, and $k$ is the wave vector, determining the sign of the current. This orbital current flowing in the $z$-direction (normal to the interface) can interact with a ferromagnetic material and exchange angular momentum. This interaction with the oscillating magnetization of the magnetic layer generates a dc component of the orbital current that can then be reconverted into a dc-charge current via IOHE (oscillating parts are also present, but not detected). The resulting oscillating orbital current, $\mathbf{j}_{L}$, depends on the current from Eq.\ \ref{eq:jl0} and is proportional to the torque exerted into the magnetization in the ferromagnet as suggested by \cite{TheorySpinHallMR}


\[
\mathbf{j}_{L} = j_{L,0}^{y} [ c_1 \hat{y} + c_2 \mathbf{m} \times (\mathbf{m} \times \hat{y}) + c_3 \mathbf{m} \times \hat{y} ]
\]

\noindent where $c_i$ are the interface-dependent coefficients, which include the orbital mixing conductance $g_o$ (interface transparency). Using the equation of magnetoelastically-driven magnetization, the generated voltage transverse to the SAWs, $V_{xy}$, is expected to be proportional to $c_2$ with an angular symmetry of $\sin(4\varphi_H)$ (see Supplementary Note 4). Additional rectification mechanisms can introduce offset voltages through electromagnetic cross-talk between the IDT and the magnetic stripe. Electromagnetic waves (EMWs) radiated by the IDTs reach the device with negligible delay and generate spurious currents that can mix with the delayed SAW-driven response. The resulting interference gives rise to a phase-dependent rectified offset voltage that does not depend on the magnetic-field polarity.

As discussed in Supplementary Note 2, the $\sin(4\varphi_H)$ term grows near the IDT resonance, whereas the offset exhibits oscillations consistent with phase-dependent cross-talk. We therefore define the AOHE contribution as the even voltage after subtracting this offset: 
$V_{xy}^{\rm AOHE} = V_{xy}^{\rm even}-V_{\rm offset}$.

\subsection*{Acoustic pumping (AP)}
Conversely, the odd voltage component reflects angular-momentum pumping from the ferromagnet, thus $V_{xy}^{\rm AP} = V_{xy}^{\rm odd}$. Both orbital and spin pumping originate from the magnetization dynamics and scale with the time derivative of the magnetization; because these dynamics are driven by the magnetoelastic interaction, we refer to the process collectively as acoustic pumping (AP). Whether AP yields spin or orbital currents depends on the intrinsic properties of the ferromagnet: Ni, with its relatively strong spin-orbit coupling, can pump both orbital and spin angular momentum, whereas Co is expected to pump predominantly spin. When a non-magnetic layer is attached, these pumped currents can be converted into a measurable charge signal through the ISHE or the IOHE. In our study, we employ light metals such as Ti and Cr, which are predicted to exhibit weak ISHE but strong IOHE \cite{OHESHE_conductivities}. The voltage arising from orbital pumping follows the angular dependence of the magnetization precession amplitude, proportional to $\mathbf{m}\times \dot{\mathbf{m}}$, and its sign reflects the SHE/OHE of the adjacent material. Accordingly, the transversal voltage associated to the acoustic pumping follows a $[\sin^2(2\varphi_H)+\sin^2(\varphi_H)]\cos(\varphi_H)$ dependence (see Supplementary Note 5).

\subsection*{Comparison across different FM$|$LM}
\begin{figure}[ht]
    \centering
    \includegraphics[width=0.6\textwidth]{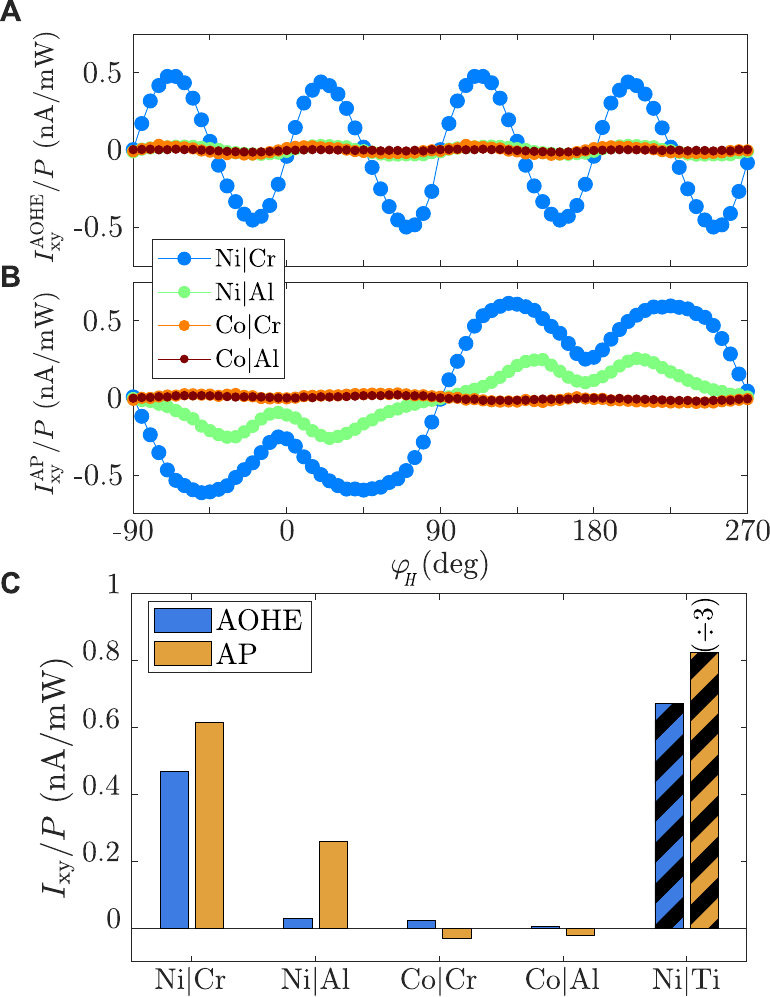}
    \caption{
    (\textbf{A}) Normalized AOHE signal for each sample as a function of the angle between the SAWs and the magnetic field at a fixed magnetic field of $\mu_0H = 35$ mT.
    (\textbf{B}) Normalized AP signal for each sample as a function of the angle between the SAWs and the magnetic field at a fixed magnetic field of $\mu_0H = 3.5$ mT for Ni-based samples and $\mu_0H = 1.3$ mT for Co-based samples.
    (\textbf{C}) Summary of the results, where the blue bars represent the AOHE signal and the gold bars represent the AP signal. For comparison, we include Ni$|$Ti measured in stripped bars; however, this sample was not part of the same batch and has a different geometry.
    }
    \label{fig:even_odd}
\end{figure}

To compare signals across different heterostructures, we convert the measured voltages into effective electrical currents, dividing the voltage signals by the device resistance and normalizing by the rf power applied to the IDT. Figure~\ref{fig:even_odd}a-b displays the angular dependence of the AOHE (measured at 35~mT) and AP (measured at 3.5~mT) for all samples. The AOHE signal $I_{xy}^{\rm AOHE}$ [Fig.~\ref{fig:even_odd}a] shows the expected $\sin(4\varphi_H)$ behavior in every device, with a large response in Ni$|$Cr, much smaller amplitudes in Ni$|$Al and Co$|$Cr ($\sim$12-15$\times$ smaller), and an essentially negligible signal in Co$|$Al ($>$40$\times$ smaller). This suggests that Cr has a good phonon orbital coupling and the generated orbital currents are only modulated in Ni and not in Co.

The acoustic pumping signal $I_{xy}^{\rm AP}$ [Fig.~\ref{fig:even_odd}b] follows the characteristic $\sin^{2}(2\varphi_H)\cos\varphi_H$ dependence, reflecting the angular variation of magnetoacoustic-wave excitation. Ni-based bilayers generate the largest AP signals, with Ni$|$Cr exceeding Ni$|$Al by approximately a factor of three, consistent with strong orbital-to-charge conversion in Cr and suggesting that the Ni/Al interface may exhibit a significant charge conversion efficiency. In contrast, Co-based samples show only weak AP responses and, notably, exhibit an opposite sign, in agreement with Co pumping predominantly spin (rather than orbital) angular momentum combined with the negative spin Hall angle of Cr \cite{leeEfficientConversionOrbital2021c}.

Figure~\ref{fig:even_odd}c compares the AOHE (blue) and AP (gold) amplitudes for all studied samples, emphasizing the reversed AP polarity between Ni- and Co-based devices and the dominant AOHE response in structures capped with Cr. For completeness and to compare with recently published results \cite{acousticOHE2025}, we also include a Ni$|$Ti device from a different fabrication batch. The Ni$|$Ti bilayer exhibits a strong AP signal---approximately three times larger than that of Ni$|$Cr---while its AOHE amplitude remains comparable to that of the Ni$|$Cr sample.

\subsection*{Power and frequency dependence}
To analyze how the AOHE and AP signals scale with the acoustic driving strength, we characterized their dependence on the input power of a single IDT. Figure~\ref{fig:2IDT}a shows the measured voltages in the Ni$|$Cr device, where the gold dots correspond to the AP signal and the blue dots to the AOHE. Both quantities are plotted as a function of the rf power (0–50 mW) applied to the IDT. The AOHE increases linearly with power, reflecting the linear growth of the phonon-driven orbital current density. In contrast, the AP signal begins to saturate at higher powers, consistent with a possible saturation of the magnetization precession amplitude \cite{casals2020generation}. Based on this behavior, we have not exceeded a maximum input power of 34 mW in the experiments presented. We also measured the power dependence as a function of frequency, shown in Fig.~\ref{fig:2IDT}b (see Supplementary Note 6), and found that the AP signal increases more rapidly than the AOHE, even though both are expected to scale similarly with frequency.

\subsection*{Vorticity dependence}
To investigate the AOHE's dependence on the vorticity of the lattice rotation, we performed an experiment on the same Ni$|$Cr sample using both IDTs simultaneously. By controlling the power ratio between the two transducers, we continuously tune the net lattice motion from purely right-hand-side propagation ( $+k$ red curve) to purely left-hand-side propagation ($-k$ blue curve), with a standing-wave configuration at equal power---where the local rotational motion is strongly suppressed\cite{StandingSAW2019}. The two IDTs were frequency-locked and driven synchronously while the total rf power was kept constant, ensuring that only the vorticity of the lattice motion was varied.

\begin{figure}[ht]
    \centering
    \includegraphics[width=0.6\textwidth]{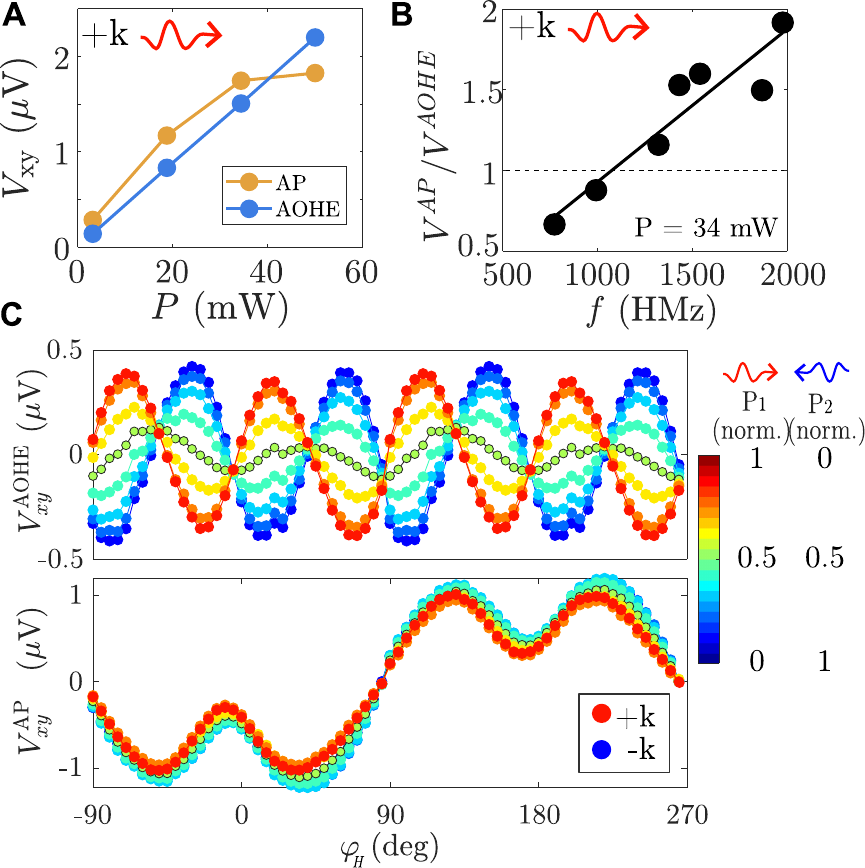}
    \caption{
    (\textbf{A}) Plot of the voltage $V_{xy}$ for each contribution: pumping (gold) and AOHE (blue) for a single propagating SAW in the +k direction. 
    (\textbf{B}) Quotient $V^{\rm AP}/V^{\rm AOHE}$ as a function of frequency at a fixed power $P= 34$ mW for a single propagating SAW in the +k direction.
    (\textbf{C}) Voltage signals with both IDTs emitting SAWs at different power ratios, but keeping the total power constant. The top panel shows the AOHE voltage $V_{xy}^{\rm AOHE}$ for different power ratios of IDTs; the color code represents each power ratio, where blue is maximum power in IDT2 and no power on IDT1. At the same time, red is maximum power in IDT1 and no power in IDT2. In green dots with black circles, we show the result of having both IDTs at the same power, which should have a dominant standing wave with some propagating. The bottom panel shows the pumping signal for the different power ratios.}
    \label{fig:2IDT}
\end{figure}

We then measured the transverse voltage as a function of the magnetic field strength and angle for different power ratios. Figure \ref{fig:2IDT}c shows the resulting curves for each contribution, AOHE (top) and AP (bottom). The color coding describes the relative power ratio between IDTs. Since the total power and frequency are kept constant, we expect minor variations in the AP signal. However, the AOHE shows a sign inversion when switching the propagation direction of SAW as shown earlier in Fig.\ \ref{fig:orbitals} and it weakens with standing waves, where the rotation of the lattice vanishes completely at the SAW nodes and becomes smaller in almost all other SAW points. We note that the curve corresponding to the pure standing-wave condition (green-filled circles) still exhibits a small oscillation. This arises because the measuring stripe is 10 $\mu$m long while the SAW wavelength at 1.3 GHz is about 3 $\mu$m, so the stripe spans a non-integer number of acoustic cycles, resulting in incomplete cancellation of the local lattice rotation (See Supplementary note 4).

\section*{Discussion}
The magnitude of the detected voltage is determined by interface properties and the efficiency with which the two layers exchange angular momentum. Ni has a comparatively large intrinsic spin-orbital coupling ($\eta_{L\cdot S}^{Ni} > 0$) and weaker orbital quenching, whereas cobalt has a much smaller coupling ($\eta_{L\cdot S}^{Co} \sim 0$) \cite{SpinCurrentsFM_Amin2019,leeEfficientConversionOrbital2021c}. In addition, Ni and Cr share compatible $d$-orbital symmetry, which may promote orbital hybridization and enhance interfacial transparency for orbital currents \cite{lyalinInterfaceTransparencyOrbital2024}.  Regarding the non-magnetic layers, Cr is predicted to exhibit a large and positive orbital Hall conductivity ($\sigma_{\rm OH}^{\rm Cr} > 0$) \cite{Salemi_PRM2022} and a small spin Hall conductivity with opposite sign, whereas Al is expected to display negligible OHE and SHE.

As a consequence, orbital-current transfer dominates in Ni$|$Cr, while Co$|$Cr is governed primarily by spin pumping and the small, negative SHE of Cr ($\sigma_{\rm SH}^{\rm Cr} < 0$) \cite{Salemi_PRM2022}. This accounts for both the much weaker acoustic pumping signal in Co$|$Cr and its opposite sign compared with Ni$|$Cr. The Ni$|$Al bilayer still produces a sizable pumping signal (roughly one-third that of Ni$|$Cr), which we attribute to strong interfacial effects, such as Rashba, and Ni's high efficiency in generating interfacial angular momentum. Recent studies have shown that even light metals as Al can have a sizable---but an order of magnitude smaller than Cr--OHE and SHE \cite{joGiganticIntrinsicOrbital2018a}.  

The amplitude of the AOHE reflects three coupled ingredients. The first is the injected orbital current density, $j_{L,0}^{y}$, set by the phonon-orbital coupling $\chi_{po}$. 
The second is the interaction of this orbital current with the adjacent ferromagnet, either through direct coupling to the interfacial orbital accumulation or via the spin channel. 
In Ni-based bilayers, the large spin-orbit coupling $\eta_{L\cdot S}^{\rm Ni}$ allows both pathways, whereas Co-based samples---with their much smaller $\eta_{L\cdot S}^{\rm Co}$---can couple only weakly and primarily through the spin channel. 
The third ingredient is the final orbital-to-charge conversion through the IOHE of the capping metal, which is strong in Cr and Ti but negligible in Al. Taken together, these factors account for the material trends observed experimentally: Ni$|$Cr exhibits a strong AOHE because it combines a large orbital Hall conductivity ($\sigma_{\rm OH}^{\rm Cr}$) with efficient interfacial coupling, whereas Ni$|$Al remains weak due to Al’s negligible OHE. Co-based samples show only small responses regardless of the cap, reflecting the limited spin-orbit coupling in Co. Finally, although Ni$|$Ti and Ni$|$Cr show comparable AOHE amplitudes, the much larger AP signal in Ni$|$Ti suggests that Cr may host a stronger phonon–orbital interaction than Ti.


We carried out additional orbital-transport measurements on the same samples---without SAW excitation---to benchmark the electrically generated signals (Supplementary Notes 7 and 8). Second-harmonic (SH) Hall and spin-transfer torque ferromagnetic resonance (STT-FMR) techniques both quantify how efficiently electrically driven spin or orbital currents exert torques on a ferromagnet. SH experiments show a weak damping-like contribution to torques in the Ni-based samples, whereas the signal falls below the resolution of the technique in the Co ones. Likewise, STT-FMR experiments show a small difference between capping materials and weak damping-like contributions in the Ni-based samples. Additionally, Ni$|$Ti bilayer exhibit a sizable response in SH measurements, consistent with  a detectable electrically driven OHE \cite{Hayashi2024_IOHE_orbital_pumping,APL_2harmonic_NiTi_2025,sunDeterminationOrbitalRelaxation2025a}. These results confirm that electrically generated orbital currents in our Cr-based systems are intrinsically weak, in agreement with earlier studies \cite{prapp2023_nicr_sttfmr}. In sharp contrast, the SAW-driven mechanisms produce large AOHE signals and strong acoustic pumping in both Ni$|$Cr and Ni$|$Ti. Moreover, while the AP signal in Ni$|$Ti is about five times larger than in Ni$|$Cr, the AOHE amplitudes are comparable, indicating that excitation through rotational-phonons strongly enhances the mechanisms responsible for generating, transferring, and coupling orbital angular momentum---far beyond what is achievable with purely electrical driving.

%

\subsection*{Conclusion}

We have demonstrated that chiral surface acoustic waves generate strong orbital currents in light-metal/ferromagnet heterostructures via both the acoustic orbital Hall effect and acoustic orbital pumping. By decomposing the SAW-driven voltages into symmetric and antisymmetric components with respect to the magnetic field, we identify two distinct mechanisms. The magnetic-field–even signal reverses with the SAW propagation direction and corresponds to the acoustic orbital Hall effect \cite{acousticOHE2025}, which originates from lattice rotation in the non-magnetic layer. In contrast, the magnetic-field–odd signal is insensitive to the propagation direction and arises from acoustic pumping driven by angular-momentum transfer in the ferromagnet. Ni$|$Cr and Ni$|$Ti bilayers exhibit large AOHE and AP signals, whereas Ni$|$Al and Co-based devices show only weak responses. These trends reflect the combined influence of phonon–orbital coupling, orbital Hall conductivity, and interfacial orbital transparency. Experiments varying SAW propagation direction confirm the orbital origin of the AOHE, revealing both magnitude suppression under standing-wave conditions and full signal inversion upon reversing SAW propagation direction.

Complementary measurements using electric fields, such as STT-FMR and second-harmonic measurements, reveal extremely weak orbital currents in the same samples, suggesting the efficiency of SAW-driven mechanisms is much better. Together, our results establish rotational-phonon excitation as a highly effective means of generating, modulating, and detecting orbital angular momentum in solids---laying the foundation for phononic orbitronics in which lattice vibrations, rather than charge currents, serve as the primary source of orbital control and information transport.


\clearpage 

%
\bibliography{references} 

@PREAMBLE{
 "\providecommand{\noopsort}[1]{}" 
 # "\providecommand{\singleletter}[1]{#1}%" 
}

@article{Gambardella_prb_OHE_2022,
  title = {Giant orbital Hall effect and orbital-to-spin conversion in $3d$, $5d$, and $4f$ metallic heterostructures},
  author = {Sala, Giacomo and Gambardella, Pietro},
  journal = {Phys. Rev. Res.},
  volume = {4},
  issue = {3},
  pages = {033037},
  numpages = {14},
  year = {2022},
  month = {Jul},
  publisher = {American Physical Society},
  doi = {10.1103/PhysRevResearch.4.033037},
  url = {https://link.aps.org/doi/10.1103/PhysRevResearch.4.033037}
}

@article{Hayashi2024_IOHE_orbital_pumping,
  author       = {Hiroki Hayashi and Dongwook Go and Satoshi Haku and Yuriy Mokrousov and Kazuya Ando},
  title        = {Observation of orbital pumping},
  journal      = {Nature Electronics},
  year         = {2024},
  month        = aug,
  volume       = {7},
  number       = {8},
  pages        = {646--652},
  doi          = {10.1038/s41928-024-01193-1},
  url          = {https://doi.org/10.1038/s41928-024-01193-1},
  issn         = {2520-1131},
  note         = {Published 2024/08/01}
}

@article{Choi2023,
author = {Choi, Young-Gwan and Jo, Daegeun and Ko, Kyung-Hun and Go, Dongwook and Kim, Kyung-Han and Park, Hee Gyum and Kim, Changyoung and Min, Byoung-Chul and Choi, Gyung-Min and Lee, Hyun-Woo},
doi = {10.1038/s41586-023-06101-9},
issn = {1476-4687},
journal = {Nature},
number = {7968},
pages = {52--56},
title = {{Observation of the orbital Hall effect in a light metal Ti}},
url = {https://doi.org/10.1038/s41586-023-06101-9},
volume = {619},
year = {2023}
}

@article{Rovillain_CoPtSAW,
  title = {Nonsymmetric spin pumping in a multiferroic heterostructure},
  author = {Rovillain, Pauline and de Oliveira, Ronei Cardoso and Marangolo, Massimiliano and Duquesne, Jean-Yves},
  journal = {Phys. Rev. B},
  volume = {102},
  issue = {18},
  pages = {184409},
  numpages = {6},
  year = {2020},
  month = {Nov},
  publisher = {American Physical Society},
  doi = {10.1103/PhysRevB.102.184409},
  url = {https://link.aps.org/doi/10.1103/PhysRevB.102.184409}
}

@article{Weiler_CoPt,
  title = {Spin Pumping with Coherent Elastic Waves},
  author = {Weiler, M. and Huebl, H. and Goerg, F. S. and Czeschka, F. D. and Gross, R. and Goennenwein, S. T. B.},
  journal = {Phys. Rev. Lett.},
  volume = {108},
  issue = {17},
  pages = {176601},
  numpages = {5},
  year = {2012},
  month = {Apr},
  publisher = {American Physical Society},
  doi = {10.1103/PhysRevLett.108.176601},
  url = {https://link.aps.org/doi/10.1103/PhysRevLett.108.176601}
}

@article{edelstein_saw_Bi_prb,
  title = {Inverse Edelstein effect induced by magnon-phonon coupling},
  author = {Xu, Mingran and Puebla, Jorge and Auvray, Florent and Rana, Bivas and Kondou, Kouta and Otani, Yoshichika},
  journal = {Phys. Rev. B},
  volume = {97},
  issue = {18},
  pages = {180301},
  numpages = {4},
  year = {2018},
  month = {May},
  publisher = {American Physical Society},
  doi = {10.1103/PhysRevB.97.180301},
  url = {https://link.aps.org/doi/10.1103/PhysRevB.97.180301}
}

@article{Puebla_2020_SAW_BiO,
doi = {10.1088/1361-6463/ab7efe},
url = {https://dx.doi.org/10.1088/1361-6463/ab7efe},
year = {2020},
month = {apr},
publisher = {IOP Publishing},
volume = {53},
number = {26},
pages = {264002},
author = {Puebla, Jorge and Xu, Mingran and Rana, Bivas and Yamamoto, Kei and Maekawa, Sadamichi and Otani, Yoshichika},
title = {Acoustic ferromagnetic resonance and spin pumping induced by surface acoustic waves},
journal = {Journal of Physics D: Applied Physics}
}

@article{prb_spinHall_3dmetals_2014,
  title = {Systematic variation of spin-orbit coupling with $d$-orbital filling: Large inverse spin Hall effect in $3d$ transition metals},
  author = {Du, Chunhui and Wang, Hailong and Yang, Fengyuan and Hammel, P. Chris},
  journal = {Phys. Rev. B},
  volume = {90},
  issue = {14},
  pages = {140407},
  numpages = {5},
  year = {2014},
  month = {Oct},
  publisher = {American Physical Society},
  doi = {10.1103/PhysRevB.90.140407},
  url = {https://link.aps.org/doi/10.1103/PhysRevB.90.140407}
}

@article{PRL2022_MR_rashba_orbitals,
  title = {Observation of the Orbital Rashba-Edelstein Magnetoresistance},
  author = {Ding, Shilei and Liang, Zhongyu and Go, Dongwook and Yun, Chao and Xue, Mingzhu and Liu, Zhou and Becker, Sven and Yang, Wenyun and Du, Honglin and Wang, Changsheng and Yang, Yingchang and Jakob, Gerhard and Kl\"aui, Mathias and Mokrousov, Yuriy and Yang, Jinbo},
  journal = {Phys. Rev. Lett.},
  volume = {128},
  issue = {6},
  pages = {067201},
  numpages = {6},
  year = {2022},
  month = {Feb},
  publisher = {American Physical Society},
  doi = {10.1103/PhysRevLett.128.067201},
  url = {https://link.aps.org/doi/10.1103/PhysRevLett.128.067201}
}

@article{Acoustic_spin_Hall_SO_saw,
author = {Takuya Kawada  and Masashi Kawaguchi  and Takumi Funato  and Hiroshi Kohno  and Masamitsu Hayashi },
title = {Acoustic spin Hall effect in strong spin-orbit metals},
journal = {Science Advances},
volume = {7},
number = {2},
pages = {eabd9697},
year = {2021},
doi = {10.1126/sciadv.abd9697},
URL = {https://www.science.org/doi/abs/10.1126/sciadv.abd9697},
eprint = {https://www.science.org/doi/pdf/10.1126/sciadv.abd9697}
}

@article{Ni_Ta_SAW_spin_rotation,
    author = {Mi, Shuai and Zhao, Chenbo and Liu, Meihong and Wang, Jianbo and Liu, Qingfang},
    title = {Detection of spin current generated by the acoustic spin–rotation coupling mechanism via acoustic voltage},
    journal = {Applied Physics Letters},
    volume = {125},
    number = {25},
    pages = {252402},
    year = {2024},
    month = {12},
    issn = {0003-6951},
    doi = {10.1063/5.0242879},
    url = {https://doi.org/10.1063/5.0242879}
}

@article{acousticOHE2025,
  author    = {Taniguchi, Mari and Haku, Satoshi and Lee, Hyun-Woo and Ando, Kazuya},
  title     = {Acoustic generation of orbital currents},
  journal   = {Nature Communications},
  year      = {2025},
  volume    = {16},
  number    = {1},
  pages     = {8038},
  doi       = {10.1038/s41467-025-62703-z},
  url       = {https://doi.org/10.1038/s41467-025-62703-z},
  issn      = {2041-1723}
}

@article{Spin_rotation_W_Pt_SAW,
  author       = {Yang Cao and Hao Ding and Yalu Zuo and Xiling Li and Yibing Zhao and Tong Li and Na Lei and Jiangwei Cao and Mingsu Si and Li Xi and Chenglong Jia and Desheng Xue and Dezheng Yang},
  title        = {Acoustic spin rotation in heavy-metal-ferromagnet bilayers},
  journal      = {Nature Communications},
  year         = {2024},
  month        = feb,
  volume       = {15},
  number       = {1},
  pages        = {1013},
  doi          = {10.1038/s41467-024-45317-9},
  url          = {https://doi.org/10.1038/s41467-024-45317-9},
  issn         = {2041-1723},
  note         = {Published 2024/02/03}
}

@article{Hayashi2023_st_fmr_niTi,
  author       = {Hiroki Hayashi and Daegeun Jo and Dongwook Go and Tenghua Gao and Satoshi Haku and Yuriy Mokrousov and Hyun-Woo Lee and Kazuya Ando},
  title        = {Observation of long-range orbital transport and giant orbital torque},
  journal      = {Communications Physics},
  year         = {2023},
  month        = feb,
  volume       = {6},
  number       = {1},
  pages        = {32},
  doi          = {10.1038/s42005-023-01139-7},
  url          = {https://doi.org/10.1038/s42005-023-01139-7},
  issn         = {2399-3650},
  note         = {Published 2023/02/06}
}

@article{McCord_AEM_2022,
author = {Müller, Cai and Durdaut, Phillip and Holländer, Rasmus B. and Kittmann, Anne and Schell, Viktor and Meyners, Dirk and Höft, Michael and Quandt, Eckhard and McCord, Jeffrey},
title = {Imaging of Love Waves and Their Interaction with Magnetic Domain Walls in Magnetoelectric Magnetic Field Sensors},
journal = {Advanced Electronic Materials},
volume = {8},
number = {6},
pages = {2200033},
doi = {https://doi.org/10.1002/aelm.202200033},
url = {https://onlinelibrary.wiley.com/doi/abs/10.1002/aelm.202200033},
year = {2022}
}

@article{SAW_electrical_adv_mat2023,
author = {Chen, Chong and Han, Lei and Liu, Peisen and Zhang, Yichi and Liang, Shixuan and Zhou, Yongjian and Zhu, Wenxuan and Fu, Sulei and Pan, Feng and Song, Cheng},
title = {Direct-Current Electrical Detection of Surface-Acoustic-Wave-Driven Ferromagnetic Resonance},
journal = {Advanced Materials},
volume = {35},
number = {38},
pages = {2302454},
doi = {https://doi.org/10.1002/adma.202302454},
url = {https://advanced.onlinelibrary.wiley.com/doi/abs/10.1002/adma.202302454},
eprint = {https://advanced.onlinelibrary.wiley.com/doi/pdf/10.1002/adma.202302454},
year = {2023}
}

@article{Labanowski2016,
	author = {Labanowski,D.  and Jung,A.  and Salahuddin,S. },
	title = {Power absorption in acoustically driven ferromagnetic resonance},
	journal = {Applied Physics Letters},
	volume = {108},
	number = {2},
	pages = {022905},
	year = {2016},
	doi = {10.1063/1.4939914},
	
	URL = { 
	https://doi.org/10.1063/1.4939914
	
	}
	
}

@article{weiler2011elastically,
  title={Elastically driven ferromagnetic resonance in nickel thin films},
  author={Weiler, Mathias and Dreher, Lothar and Heeg, Christian and Huebl, Hans and Gross, Rudolf and Brandt, Martin S and Goennenwein, Sebastian TJ},
  journal={Physical Review Letters},
  volume={106},
  number={11},
  pages={117601},
  year={2011},
  publisher={APS},
  doi={10.1103/PhysRevLett.106.117601}
}

@article{casals2020generation,
  title={Generation and Imaging of Magnetoacoustic Waves over Millimeter Distances},
  author={Casals, B and Statuto, N and Foerster, M and Hern{\'a}ndez-M{\'i}nguez, A and Cichelero, R and Manshausen, P and Mandziak, A and Aballe, L and Hern{\`a}ndez, J M and Maci{\`a}, F},
  journal={Physical Review Letters},
  volume={124},
  number={13},
  pages={137202},
  year={2020},
  publisher={APS},
  doi={10.1103/PhysRevLett.124.137202}
}

@article{SAW_Co_Ni_Rovirola,
  title = {Study of the magnetoelastic effect in nickel and cobalt thin films at GHz range using x-ray microscopy},
  author = {Rovirola, Marc and Waqas Khaliq, M. and Gustafson, Travis and Sosa-Barth, Fiona and Casals, Blai and Hern\`andez, Joan Manel and Ruiz-G\'omez, Sandra and Ni\~no, Miguel Angel and Aballe, Luc\'{\i}a and Hern\'andez-M\'{\i}nguez, Alberto and Foerster, Michael and Maci\`a, Ferran},
  journal = {Phys. Rev. Res.},
  volume = {6},
  issue = {2},
  pages = {023285},
  numpages = {9},
  year = {2024},
  month = {Jun},
  publisher = {American Physical Society},
  doi = {10.1103/PhysRevResearch.6.023285},
  url = {https://link.aps.org/doi/10.1103/PhysRevResearch.6.023285}
}

@article{prapp2023_nicr_sttfmr,
  title = {Sign Reversal of Fieldlike Spin-Orbit Torque in an Ultrathin $\mathrm{Cr}/\mathrm{Ni}$ Bilayer},
  author = {Bose, Arnab and Singh, Hanuman and Kushwaha, Varun Kumar and Bhuktare, Swapnil and Dutta, Sutapa and Tulapurkar, Ashwin A.},
  journal = {Phys. Rev. Appl.},
  volume = {9},
  issue = {1},
  pages = {014022},
  numpages = {8},
  year = {2018},
  month = {Jan},
  publisher = {American Physical Society},
  doi = {10.1103/PhysRevApplied.9.014022},
  url = {https://link.aps.org/doi/10.1103/PhysRevApplied.9.014022}
}

@article{APL_2harmonic_NiTi_2025,
    author = {Mahapatra, Dhananjaya and Bhunia, Harekrishna and Miah, Abu Bakkar and Aon, Soumik and Mitra, Partha},
    title = {Evidence of orbital Hall current-induced correlation in second-harmonic response of longitudinal and transverse voltage in light metal–ferromagnet bilayers},
    journal = {Applied Physics Letters},
    volume = {126},
    number = {24},
    pages = {242406},
    year = {2025},
    month = {06},
    issn = {0003-6951},
    doi = {10.1063/5.0263240},
    url = {https://doi.org/10.1063/5.0263240}
}

@article{Salemi_PRM2022,
  title = {First-principles theory of intrinsic spin and orbital Hall and Nernst effects in metallic monoatomic crystals},
  author = {Salemi, Leandro and Oppeneer, Peter M.},
  journal = {Phys. Rev. Mater.},
  volume = {6},
  issue = {9},
  pages = {095001},
  numpages = {12},
  year = {2022},
  month = {Sep},
  publisher = {American Physical Society},
  doi = {10.1103/PhysRevMaterials.6.095001},
  url = {https://link.aps.org/doi/10.1103/PhysRevMaterials.6.095001}
}

@article{OHE_Texture_prl2018,
  title = {Intrinsic Spin and Orbital Hall Effects from Orbital Texture},
  author = {Go, Dongwook and Jo, Daegeun and Kim, Changyoung and Lee, Hyun-Woo},
  journal = {Phys. Rev. Lett.},
  volume = {121},
  issue = {8},
  pages = {086602},
  numpages = {6},
  year = {2018},
  month = {Aug},
  publisher = {American Physical Society},
  doi = {10.1103/PhysRevLett.121.086602},
  url = {https://link.aps.org/doi/10.1103/PhysRevLett.121.086602}
}

@article{StandingSAW2019,
author = "Foerster, Michael and Statuto, Nahuel and Casals, Blai and Hern{\'{a}}ndez-M{\'\i}nguez, Alberto and Finizio, Simone and Mandziak, Ania and Aballe, Lucia and Hern{\`{a}}ndez Ferr{\`{a}}s, Joan Manel and Maci{\`{a}}, Ferran",
title = "{Quantification of propagating and standing surface acoustic waves by stroboscopic X-ray photoemission electron microscopy}",
journal = "Journal of Synchrotron Radiation",
year = "2019",
volume = "26",
number = "1",
pages = "184--193",
month = "Jan",
doi = {10.1107/S1600577518015370},
url = {https://doi.org/10.1107/S1600577518015370}
}

@article{yang2021acoustic,
  title={Acoustic control of magnetism toward energy-efficient applications},
  author={Yang, Wei-Gang and Schmidt, Holger},
  journal={Applied Physics Reviews},
  volume={8},
  number={2},
  pages={021304},
  year={2021},
  publisher={AIP Publishing},
  doi={10.1063/5.0042138}
}

@article{rovirola2023_physrevapp,
  title = {Resonant and Off-Resonant Magnetoacoustic Waves in Epitaxial {{Fe$_3$Si/GaAs}} Hybrid Structures},
  author = {Rovirola, Marc and Waqas Khaliq, M. and Casals, Blai and Foerster, Michael and Ni\~no, Miguel Angel and Aballe, Luc\'{\i}a and Herfort, Jens and Hern\`andez, Joan Manel and Maci\`a, Ferran and Hern\'andez-M\'{\i}nguez, Alberto},
  journal = {Physics Review Applied},
  volume = {20},
  issue = {3},
  pages = {034052},
  numpages = {9},
  year = {2023},
  month = {Sep},
  publisher = {American Physical Society},
  doi = {10.1103/PhysRevApplied.20.034052},
  url = {https://link.aps.org/doi/10.1103/PhysRevApplied.20.034052}
}

@article{puebla2022perspectives,
  title={Perspectives on spintronics with surface acoustic waves},
  author={Puebla, Jon and Hwang, Youngho and Maekawa, Sadamichi and Otani, Yoshichika},
  journal={Applied Physics Letters},
  volume={120},
  number={22},
  pages={220502},
  year={2022},
  publisher={AIP Publishing},
  doi={10.1063/5.0093654}
}

@article{gowthamTravelingSurfaceSpinwave2015b,
  title = {Traveling Surface Spin-Wave Resonance Spectroscopy Using Surface Acoustic Waves},
  author = {Gowtham, P. G. and Moriyama, T. and Ralph, D. C. and Buhrman, R. A.},
  year = {2015},
  month = dec,
  journal = {Journal of Applied Physics},
  volume = {118},
  number = {23},
  pages = {233910},
  issn = {0021-8979, 1089-7550},
  doi = {10.1063/1.4938390},
  langid = {english},
}

@article{Alberto_2020,
  title = {Large Nonreciprocal Propagation of Surface Acoustic Waves in Epitaxial Ferromagnetic/Semiconductor Hybrid Structures},
  author = {Hern\'andez-M\'{\i}nguez, A. and Maci\`a, F. and Hern\`andez, J. M. and Herfort, J. and Santos, P. V.},
  journal = {Physics Review Applied},
  volume = {13},
  issue = {4},
  pages = {044018},
  numpages = {14},
  year = {2020},
  month = {Apr},
  publisher = {American Physical Society},
  doi = {10.1103/PhysRevApplied.13.044018},
  url = {https://link.aps.org/doi/10.1103/PhysRevApplied.13.044018}
}

@article{casalsGenerationImagingMagnetoacoustic2020,
  title = {Generation and {{Imaging}} of {{Magnetoacoustic Waves}} over {{Millimeter Distances}}},
  author = {Casals, Blai and Statuto, Nahuel and Foerster, Michael and {Hern{\'a}ndez-M{\'i}nguez}, Alberto and Cichelero, Rafael and Manshausen, Peter and Mandziak, Ania and Aballe, Luc{\'i}a and Hern{\`a}ndez, Joan Manel and Maci{\`a}, Ferran},
  year = {2020},
  month = apr,
  journal = {Physics Review Letters},
  volume = {124},
  number = {13},
  pages = {137202},
  issn = {0031-9007, 1079-7114},
  doi = {10.1103/PhysRevLett.124.137202},
  langid = {english}
}

@article{Xu_magnetorotation_2020,
author = {Mingran Xu  and Kei Yamamoto  and Jorge Puebla  and Korbinian Baumgaertl  and Bivas Rana  and Katsuya Miura  and Hiromasa Takahashi  and Dirk Grundler  and Sadamichi Maekawa  and Yoshichika Otani },
title = {Nonreciprocal surface acoustic wave propagation via magneto-rotation coupling},
journal = {Science Advances},
volume = {6},
number = {32},
pages = {eabb1724},
year = {2020},
doi = {10.1126/sciadv.abb1724},
URL = {https://www.science.org/doi/abs/10.1126/sciadv.abb1724}
}

@article{GiantNonReciprocalShah2020,
author = {Piyush J. Shah  and Derek A. Bas  and Ivan Lisenkov  and Alexei Matyushov  and Nian X. Sun  and Michael R. Page },
title = {Giant nonreciprocity of surface acoustic waves enabled by the magnetoelastic interaction},
journal = {Science Advances},
volume = {6},
number = {49},
pages = {eabc5648},
year = {2020},
doi = {10.1126/sciadv.abc5648}
}

@article{TheorySpinHallMR,
  title = {Theory of spin Hall magnetoresistance},
  author = {Chen, Yan-Ting and Takahashi, Saburo and Nakayama, Hiroyasu and Althammer, Matthias and Goennenwein, Sebastian T. B. and Saitoh, Eiji and Bauer, Gerrit E. W.},
  journal = {Phys. Rev. B},
  volume = {87},
  issue = {14},
  pages = {144411},
  numpages = {9},
  year = {2013},
  month = {Apr},
  publisher = {American Physical Society},
  doi = {10.1103/PhysRevB.87.144411},
  url = {https://link.aps.org/doi/10.1103/PhysRevB.87.144411}
}

@article{OHESHE_conductivities,
  title = {Gigantic intrinsic orbital Hall effects in weakly spin-orbit coupled metals},
  author = {Jo, Daegeun and Go, Dongwook and Lee, Hyun-Woo},
  journal = {Phys. Rev. B},
  volume = {98},
  issue = {21},
  pages = {214405},
  numpages = {11},
  year = {2018},
  month = {Dec},
  publisher = {American Physical Society},
  doi = {10.1103/PhysRevB.98.214405},
  url = {https://link.aps.org/doi/10.1103/PhysRevB.98.214405}
}

@article{SpinCurrentsFM_Amin2019,
  title = {Intrinsic spin currents in ferromagnets},
  author = {Amin, V. P. and Li, Junwen and Stiles, M. D. and Haney, P. M.},
  journal = {Phys. Rev. B},
  volume = {99},
  issue = {22},
  pages = {220405},
  numpages = {5},
  year = {2019},
  month = {Jun},
  publisher = {American Physical Society},
  doi = {10.1103/PhysRevB.99.220405},
  url = {https://link.aps.org/doi/10.1103/PhysRevB.99.220405}
}

@article{leeEfficientConversionOrbital2021c,
  title = {Efficient Conversion of Orbital {{Hall}} Current to Spin Current for Spin-Orbit Torque Switching},
  author = {Lee, Soogil and Kang, Min-Gu and Go, Dongwook and Kim, Dohyoung and Kang, Jun-Ho and Lee, Taekhyeon and Lee, Geun-Hee and Kang, Jaimin and Lee, Nyun Jong and Mokrousov, Yuriy and Kim, Sanghoon and Kim, Kab-Jin and Lee, Kyung-Jin and Park, Byong-Guk},
  year = 2021,
  month = nov,
  journal = {Communications Physics},
  volume = {4},
  number = {1},
  pages = {234},
  issn = {2399-3650},
  doi = {10.1038/s42005-021-00737-7}
}

@article{joGiganticIntrinsicOrbital2018a,
  title = {Gigantic Intrinsic Orbital {{Hall}} Effects in Weakly Spin-Orbit Coupled Metals},
  author = {Jo, Daegeun and Go, Dongwook and Lee, Hyun-Woo},
  year = 2018,
  month = dec,
  journal = {Physical Review B},
  volume = {98},
  number = {21},
  pages = {214405},
  issn = {2469-9950, 2469-9969},
  doi = {10.1103/PhysRevB.98.214405},
  urldate = {2025-11-17},
  langid = {english}
}

@misc{wuAcousticOrbitalHall2025,
  title = {Acoustic Orbital {{Hall}} Effect and Orbital Pumping in Light-Metal-Ferromagnet Bilayers},
  author = {Wu, Mingxing and Ding, Shilei and Matsumoto, Hiroki and Gambardella, Pietro},
  year = 2025,
  month = nov,
  number = {arXiv:2511.02388},
  eprint = {2511.02388},
  primaryclass = {cond-mat},
  publisher = {arXiv},
  doi = {10.48550/arXiv.2511.02388},
  urldate = {2025-11-08},
  archiveprefix = {arXiv}
}

@article{lyalinInterfaceTransparencyOrbital2024,
  title = {Interface Transparency to Orbital Current},
  author = {Lyalin, Igor and Kawakami, Roland K.},
  year = 2024,
  month = sep,
  journal = {Physical Review B},
  volume = {110},
  number = {10},
  pages = {104418},
  issn = {2469-9950, 2469-9969},
  doi = {10.1103/PhysRevB.110.104418},
  urldate = {2025-11-19},
  langid = {english}
}

@article{sunDeterminationOrbitalRelaxation2025a,
  title = {Determination of Orbital Relaxation in {{Ti}}/{{Ni}} Heterostructure via Orbital Pumping},
  author = {Sun, Rui and Nabei, Yoji and McConnell, Aeron and Zhang, Xiaotong and Comstock, Andrew and Jones, Hana and Gyawali, Rishiram and Xiong, Yuzan and Wang, Ziqi and Liu, Jun and Zhang, Wei and Sun, Dali},
  year = 2025,
  month = sep,
  journal = {Journal of Applied Physics},
  volume = {138},
  number = {12},
  pages = {123905},
  issn = {0021-8979, 1089-7550},
  doi = {10.1063/5.0292745},
  urldate = {2025-11-24}
}

@article{martinrioSuppressionSpinRectification2022,
  title = {Suppression of Spin Rectification Effects in Spin Pumping Experiments},
  author = {{Martin-Rio}, Sergi and Frontera, Carlos and Pomar, Alberto and Balcells, Lluis and Martinez, Benjamin},
  year = 2022,
  month = jan,
  journal = {Scientific Reports},
  volume = {12},
  pages = {224},
  issn = {2045-2322},
  doi = {10.1038/s41598-021-04319-z},
  urldate = {2025-11-25},
  pmcid = {PMC8742073},
  pmid = {34997112}
}

@article{SAWunid1,
author = {K. Yamanouchi  and H. Furuyashiki },
title = {New low-loss SAW filter using internal floating electrode reflection types of single-phase unidirectional transducer},
journal = {Electronics Letters},
volume = {20},
issue = {24},
pages = {989-990},
year = {1984},
doi = {10.1049/el:19840672},
URL = {https://digital-library.theiet.org/doi/abs/10.1049/el%3A19840672},
}

@inproceedings{morganInvestigationNovelFloatingelectrode2000a,
  title = {Investigation of Novel Floating-Electrode Unidirectional {{SAW}} Transducers ({{FEUDT}}'s)},
  booktitle = {2000 {{IEEE Ultrasonics Symposium}}. {{Proceedings}}. {{An International Symposium}} ({{Cat}}. {{No}}.{{00CH37121}})},
  author = {Morgan, D.P.},
  year = 2000,
  month = oct,
  volume = {1},
  pages = {15-19 vol.1},
  issn = {1051-0117},
  doi = {10.1109/ULTSYM.2000.922497},
  urldate = {2025-11-25}
}

@article{Manchon2019,
  title = {Current-induced spin-orbit torques in ferromagnetic and antiferromagnetic systems},
  author = {Manchon, A. and \ifmmode \check{Z}\else \v{Z}\fi{}elezn\'y, J. and Miron, I. M. and Jungwirth, T. and Sinova, J. and Thiaville, A. and Garello, K. and Gambardella, P.},
  journal = {Rev. Mod. Phys.},
  volume = {91},
  issue = {3},
  pages = {035004},
  numpages = {80},
  year = {2019},
  month = {Sep},
  publisher = {American Physical Society},
  doi = {10.1103/RevModPhys.91.035004},
  url = {https://link.aps.org/doi/10.1103/RevModPhys.91.035004}
}
\bibliographystyle{aip}

%
%
%
%
%
%

\section*{Acknowledgments}
 M.R., B.C., J.M.H., J.O., Q.B., A.C. and F.M. acknowledge funding from \\ MCIN/AEI/10.13039/501100011033 through grant number: PDC2023-145910-I00 and PID2023-150721OB-I00.
 BC acknowledges funding through grant number:
 PID2022-136762NA-I00
 This project received funding from the MCIN through grant numbers CNS2024-154658, PID2024-157112OB-C53, PID2024-157112OB-C54, and PCI2025-163257, and the Fundación Ramón Areces via project CIVP22S18216.
 H.M. acknowledges support from the FPI-UAM PhD scholarship program.

\newpage


\renewcommand{\thefigure}{S\arabic{figure}}
\renewcommand{\thetable}{S\arabic{table}}
\renewcommand{\theequation}{S\arabic{equation}}
\renewcommand{\thepage}{S\arabic{page}}
\setcounter{figure}{0}
\setcounter{table}{0}
\setcounter{equation}{0}
\setcounter{page}{1} 


\begin{center}
\section*{Supplementary Materials for\\ \scititle}

Marc Rovirola$^{1,2,*}$, J\'ulia \`Odena$^{1}$, Anna Castellv\'{\i}$^{1}$,
Quim Badosa$^{1}$, Blai Casals$^{3,2}$, \and Adri\'an Gud\'{\i}n$^{4,5}$, Haripriya Madathil$^{4}$, Fernando Ajejas$^{5}$, Paolo Perna$^{5}$, \and Alberto Hern\'andez-M\'{\i}nguez$^{6}$, Joan Manel Hern\`andez$^{1,2}$, Sa\"ul V\'elez$^{4,7,\ddagger}$, \and Ferran Maci\`a$^{1,2,\dagger}$
\and \\
\small$^\ast$marc.rovirola@ub.edu \and \\
\small$^\ddagger$saul.velez@uam.es \and \\
\small$^\dagger$ferran.macia@ub.edu
\end{center}


\newpage


\subsection*{Materials and Methods}
\subsubsection*{Acoustic device properties}
The studied magnetoacoustic devices have two identical IDTs separated by 2 mm to allow for transmission ($S_{12}$, $S_{21}$) and reflection ($S_{11}$, $S_{22}$) measurements. In the acoustic path, two thin strips of the bilayer structures (FM$|$LM) are patterned perpendicular to the SAW propagation direction with gold contacts on each side to measure the transverse voltage. IDTs are placed at 0.55 and 1.45 mm from the measured strip, causing the SAWs to arrive with different strengths to the sample. The bilayer strips are 0.4 mm in length, 10 $\mu$m in width to minimize rectification effects \cite{martinrioSuppressionSpinRectification2022}, and the thickness of each FMn  and LM layer is 10 nm.

The IDTs used in the experiments are floating electrode unidirectional transducers (FEUDTs), see Refs. \cite{SAWunid1} and \cite{morganInvestigationNovelFloatingelectrode2000a}. Our design consists of four electrodes per period: a ground electrode, two signal electrodes of different widths, and a floating electrode between the two signal electrodes. This design provides both unidirectionality and excitation of even and odd harmonics. Figure \ref{setup_dim}b shows the transmission of these IDTs, where the peaks are the resonant frequencies of the IDTs, where SAWs are sent on the surface of the LiNbO$_3$.

\subsubsection*{Separation into field-symmetric (even) and field-antisymmetric (odd) voltage}
The overall transverse voltage is calculated for all angles and magnetic field strengths. This outputs a 2D matrix ($M(\theta,H)$), from which we extract even and odd contributions by copying this 2D matrix and flipping it along the magnetic field axis. The field-symmetric contribution, $V_{xy}^{FS}=V_{xy}^{\rm even}$, is the semi-sum of the matrices, while the field-antisymmetric $V_{xy}^{FA}=V_{xy}^{\rm odd}$ is the semi-difference.

\begin{equation}
    V_{xy}^{FS} = \frac{M(\theta,H)+M(\theta,-H)}{2},
\end{equation}
\begin{equation}
    V_{xy}^{FA} = \frac{M(\theta,H)-M(\theta,-H)}{2}.
\end{equation}

\subsection*{Supplementary Note 1: Setup}

Our devices have identical IDTs separated by 2 mm to allow for transmission and reflection measurements (see Fig. \ref{setup_dim} (a)). Different patterns are defined in the acoustic path, either transversal thin or longitudinal wide strips of the bilayer structures (FM$|$LM) are patterned together with gold contacts at the edges to measure both the transverse or longitudinal voltages and the SAW transmission signals.


\begin{figure}[ht]
    \centering
    \includegraphics[width=1\columnwidth]{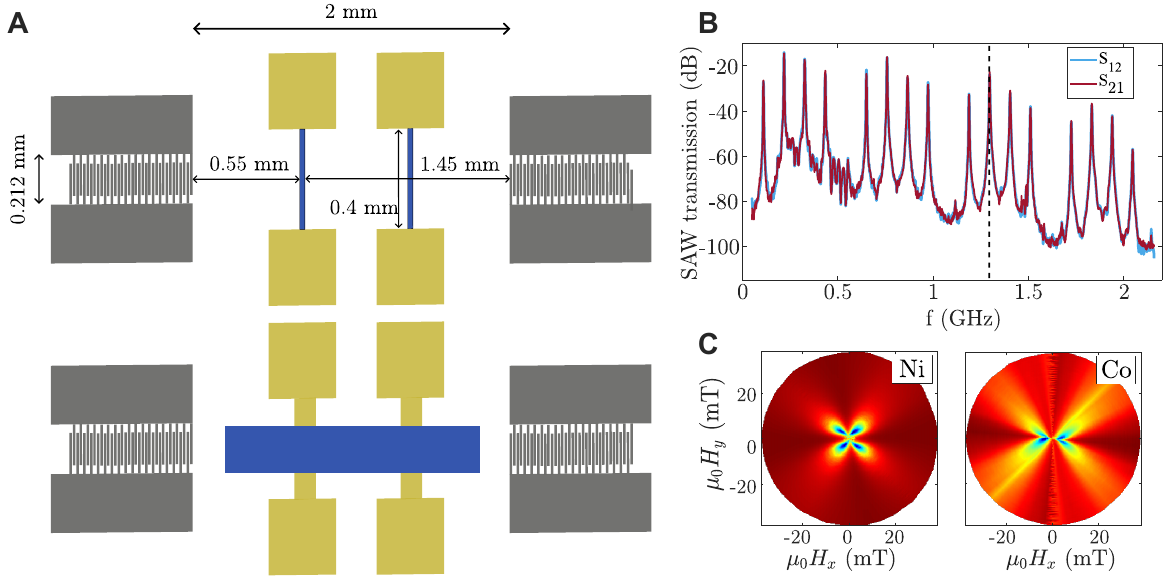}
    \caption{
    (\textbf{A}) Schematics of the magnetoacoustic device with the dimensions. The IDTs are separated by 2 mm, and in the acoustic path, we have two transversal rectangles of the material under study. The transverse bars are 10 $\mu$m in width, and the FM$|$NM are 10nm/10nm.
    (\textbf{B}) SAW transmission $S_{21}$ from 0 to 2.5 GHz. The peaks show the resonant behavior of the IDTs at which SAWs are efficiently sent. The fundamental frequency is 111 MHz. 
    (\textbf{C}) Transmission measurements of Ni-based sample (left) and Co-based sample (right) as a function of the strength and direction of the magnetic field.
    }
    \label{setup_dim}
\end{figure}


Figure~\ref{setup_dim}b shows the transmission spectrum of the IDTs, where the peaks correspond to their resonance frequencies and thus to the excitation of propagating SAWs in LiNbO$_3$. As expected from the IDT geometry, the efficiency decreases at higher harmonics, and some frequencies couple more effectively than others. We use 1.3 GHz as a representative case in the main text. However, all frequencies are measured in all samples and results can be found in Supplementary Note 6. Figure~\ref{setup_dim}c displays the transmitted signal $S_{21}$ at 1321 MHz for a Ni-based sample (left) and a Co-based sample (right), using longitudinal wide stripes positioned along the acoustic path. A time-gating technique is employed to isolate the transmitted SAW and suppress electromagnetic leakage. The blue lobes indicate transmission minima, corresponding to the angles where energy transfer from the SAW to the magnetic system is maximized. In both materials, this occurs near 45 degrees, in agreement with theoretical predictions \cite{weiler2011elastically,gowthamTravelingSurfaceSpinwave2015b,Labanowski2016}. The field dependence peaks at the magnetic-resonance condition, which differs between Ni and Co due to their distinct dispersion relations and can be approximated using Kittel’s equation

\begin{equation}
    f = \gamma \mu_0 \sqrt{(H_{\rm ext}+H_{\rm anis})(H_{\rm ext}+H_{\rm anis} + M_{\rm eff})},
\end{equation}

\noindent where $\gamma$ is approximated to 28 GHz/T, $H_{\rm ext}$ is the strength of the external magnetic field, $H_{\rm anis}$ is the anisotropy field and $M_{\rm eff}$ is the effective magnetization. The values of $\mu_0M_{\rm eff}$ can be found in Table \ref{tab:samples}. However, SAWs will alter the energy landscape and shift the resonances \cite{gowthamTravelingSurfaceSpinwave2015b}. Through this approximation and solving for $H_{ext}$ at a resonance frequency of 1321 MHz, we can find $H_{\rm res}^{Ni} \approx 3.5$ mT for Ni, and  $H_{\rm res} \approx 1.3$ mT for Co. These values are then used to extract the $V_{\rm AP}$ for each sample.

\subsection*{Supplementary Note 2: Behavior of the voltage components around the IDT resonance frequency}

\begin{figure}[htb!]
    \centering
    \includegraphics[width=1\columnwidth]{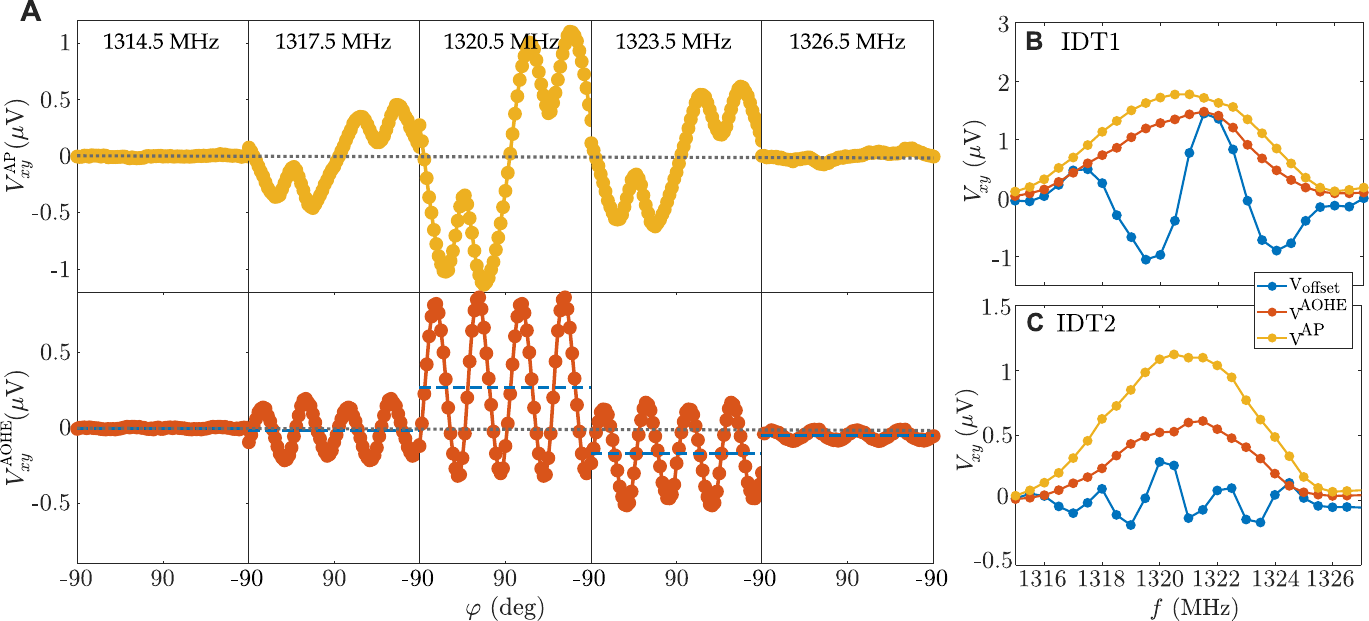}
    \caption{ 
    (\textbf{A}) Angular dependence of the field-symmetric 
    ($V_{xy}^{\rm AHOE}$, measured at $\mu_0H = 35~\text{mT}$)  
    and field-antisymmetric ($V_{xy}^{\rm AP}$, measured at $\mu_0H \approx 3.5~\text{mT}$) voltage components for frequencies within $\pm 5$ MHz of the 8th IDT harmonic (1321 MHz). Both AHOE and AP amplitudes increase near the IDT resonance, while the offset $V_{\rm offset}$ shows no systematic trend.
    (\textbf{B, C}) Maximum amplitude of each component as a function of frequency. 
    Both $V_{xy}^{\rm AHOE}$ and $V_{xy}^{\rm AP}$ peak at the IDT resonance, 
    whereas $V_{\rm offset}$ oscillates with frequency due to EMW--SAW cross-talk. 
    The oscillation period matches the phase delay $\Delta\phi = 2\pi f\, x / v_{\rm SAW}$ associated with the IDT--device separations ($x = 0.85~\text{mm}$ for IDT\,1 and $x = 1.9~\text{mm}$ for IDT\,2), confirming the EMW origin of the offset.
    }
\label{V_aroundIDtfreq}
\end{figure}

We studied how the measured voltage evolves within a $\pm 5$ MHz window around the IDT resonance, near the 8th harmonic (1321 MHz), corresponding to the data shown in Fig.~\ref{fig:even_odd} of the main text. As before, the voltage is decomposed into field-symmetric and field-antisymmetric components, associated respectively with the AHOE $V_{xy}^{\rm even}=V_{xy}^{\rm AHOE}+V_{\rm offset}$ and AP ($V_{xy}^{\rm odd}=V_{xy}^{\rm AP}$). Figure~\ref{V_aroundIDtfreq}a shows both components as a function of field angle ($V_{xy}^{\rm even}$ measured at $\mu_0H = 35$ mT and $V_{xy}^{\rm AP}$ at resonance, around $\mu_0H = 3.5$ mT). Both the AHOE and AP amplitudes increase near the IDT resonance, whereas the offset $V_{\rm offset}$ takes on seemingly random values at each frequency.

To clarify this behavior, we extract the maximum amplitude of each component at every frequency and plot the results in Figs.~\ref{V_aroundIDtfreq}b and ~\ref{V_aroundIDtfreq}c using IDT\,1 and IDT\,2. Both $V_{xy}^{\rm AHOE}$ and $V_{xy}^{\rm AP}$ peak at the IDT resonance, while $V_{\rm offset}$ exhibits a clear oscillation with frequency. The oscillation period for IDT\,1 is approximately $4.5$ MHz, consistent with the phase delay
\[
\Delta\phi = 2\pi f \frac{x}{v_{\rm SAW}},
\]
between the SAW and EMW for a separation of $x = 0.85$ mm between the IDT center and the device. For IDT\,2, located farther away ($x = 1.9$ mm), the observed oscillation period is about $2$ MHz—roughly a factor of $2.2$ smaller—as expected from the longer propagation distance. These results confirm that the offset in the symmetric voltage component originates from rectification processes associated with EMW--SAW cross-talk.


\subsection*{Supplementary Note 3: Magnetization dynamics driven by SAW}

\subsubsection*{Linearization of Landau-Lifshitz in rotated frame}

To find how the magnetization reacts to the SAWs we solve the Landau-Lifshitz equation

\begin{equation}\label{LLGfull}
\frac{d\bm m}{dt} = -\gamma\,\bm m\times\bm H_{\rm eff} 
\end{equation}

\noindent where $\gamma>0$ is the gyromagnetic ratio and the Gilbert damping will be phenomenologically added later on. The associated effective field comes from:
\[
\bm H_{\rm eff} = -\frac{1}{\mu_0 M_s}\frac{\partial F}{\partial\bm m} + \bm h_{\rm drive}(t),
\]

\noindent where $F$ are the static energy terms that can include anisotropy, Zeeman, and demagnetization, and $\bm h_{\rm drive}(t)$ is the driving field, in this case, the magnetoelastic field.

We solve the LLG in a different coordinate system ($x'$,$y'$,$z'$), where the magnetization equilibrium $m_0$ is aligned with $x'$. The new coordinate system is given by [see Fig.\ \ref{fig_supp:m_coordinates}a] 

\begin{equation}
    x' = m_0, \qquad y'\equiv \text{in-plane transverse}, \qquad z' \equiv \text{out of plane}.
\end{equation}

\begin{figure}[htb!]
    \centering
    \includegraphics[width=0.6\columnwidth]{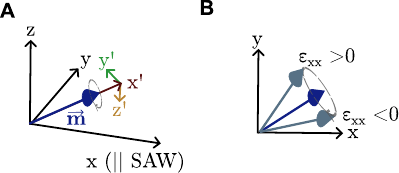}
    \caption{
    (\textbf{A}) Rotated frame of reference where $x'=m_0$ is the equilibrium direction, $y'$ is the perpendicular in-plane and $z'$ is the perpendicular out-of-plane.
    (\textbf{B}) Schematics of the magnetoelastic effect on the magnetization showing both tensile ($\varepsilon_{xx}>0$) and compressive ($\varepsilon_{xx}<0$) strain states.
    }
    \label{fig_supp:m_coordinates}
\end{figure}

In the new reference, the magnetization precesses around $x'$ and can be defined as $\delta m = (0, m_{y'},m_{z'})$ with the driving fields perpendicular to $x'$ as follows: $h_{\rm drive} = (0,h_{y'}(t),h_{z'}(t)) $. Linearizing LLG equation, keeping first order ($ \delta m_i \delta m_i \approx 0$) and assuming a oscillatory solution ($e^{-i\omega t}$) we obtain:

\begin{align}
    i\omega m_{y'} = -\gamma (H_{\rm eff} m_{z'} - h_{z'}), \\
    i\omega m_{z'} = -\gamma ( h_{y'} +H_{\rm eff} m_{y'}).
\end{align}

One can reorganize it and write it in matrix notation reading as

\begin{equation}
    \begin{pmatrix}
        m_{y'} \\ m_{z'}
    \end{pmatrix} =
    \frac{\gamma}{D(\omega)} 
    \begin{pmatrix}
        i\omega  & \gamma H_{\rm eff} \\ - \gamma H_{\rm eff} & i\omega 
    \end{pmatrix}
    \begin{pmatrix}
        h_{y'} \\ h_{z'}
    \end{pmatrix},
\end{equation}

\noindent where $D(\omega) = \omega_r^2 - \omega^2 + i\omega\Gamma$ is the determinant of the matrix $\omega_r \equiv \gamma H_{\rm eff}$, and $\Gamma \equiv 2\alpha\,\gamma H_{\rm eff} = 2\alpha\,\omega_r$ is the phenomenological damping. Next, the susceptibility can be defined as
\begin{equation}
    \chi(\omega) = \frac{\gamma}{D(\omega)} 
    \begin{pmatrix}
        i\omega & \gamma H_{\rm eff} \\ - \gamma H_{\rm eff} & i\omega
    \end{pmatrix}
\end{equation}

\noindent
and finally, the magnetization is described as
\begin{equation}\label{eq_supp:m_prime}
    \begin{pmatrix}
        m_{y'} \\ m_{z'}
    \end{pmatrix} = \chi(\omega)
    \begin{pmatrix}
        h_{y'} \\ h_{z'}
    \end{pmatrix}.
\end{equation}

\subsubsection*{Magnetoelastic driving field}
To obtain the driving field, we begin by writing the magnetoelastic energy in a phenomenological form and keeping only the terms that contribute for our geometry:

\begin{equation}
    F_{me} = \frac{B_1}{M_s}(\varepsilon_{xx}m_{x}^2+\varepsilon_{zz}m_{z}^2) + \frac{2B_2}{M_s}\varepsilon_{xz}m_xm_{z},
\end{equation}

\noindent where $B_i$ are the magnetoelastic constants that determine the magnon-phonon coupling strength, $m_i$ are the normalized magnetization components ($m = (\cos(\varphi_H),\sin(\varphi_H),0)$, and $\varepsilon_{ij}$ is the strain tensor.
The strain tensor can be determined by the atom displacement equations, which for a propagating Rayleigh SAWs are

\begin{equation} \label{eq_supp:u}
\mathbf{u} = 
\begin{pmatrix}
    u_x \\ u_y \\u_z
\end{pmatrix} = 
\begin{pmatrix}
u_0e^{kz}\cos\!\big(kx - \omega t\big) \\
0 \\
-\beta\,u_0e^{kz}\sin\!\big(kx - \omega t\big)
\end{pmatrix},
\end{equation}

\noindent where $u_0$ is the amplitude of the atom displacement, $\beta$ determines the ellipticity of the motion, $k=2\pi/\lambda$ and $\omega=2\pi f$. Then, taking the partial derivatives
\begin{equation}
\varepsilon_{ij} = \frac{1}{2} \left(\frac{\partial u_j}{\partial i} + \frac{\partial u_i}{\partial j}\right)
\end{equation}
we obtain
\begin{equation} \label{eq_sm:strain}
\begin{pmatrix}
    \varepsilon_{xx} \\
    \varepsilon_{xz} \\
    \varepsilon_{zz}
\end{pmatrix}
    =
    \begin{pmatrix}
    -k u_0e^{kz}\sin\!\big(kx - \omega t\big) \\
    \frac{1}{2} k u_0e^{kz}\cos\!\big(kx - \omega t\big) \left(1 - \beta\right) \\
    -\beta k u_0e^{kz}\sin\!\big(kx - \omega t\big)
    \end{pmatrix}.
\end{equation}

The oscillating strain goes fro tensile to compressive states driving the magnetization [see Fig.\ \ref{fig_supp:m_coordinates}b]. Then the driving effective field associated to the magnetoelastic energy is thus

\begin{equation}
\label{eq:h_me_full}
\mu_0\mathbf{h}_{\text{me}} = -\nabla_{\mathbf{m}} F_{\text{me}} =
\left\{
\begin{aligned}
\mu_0 h_{\text{me},x} &= -2\frac{B_1}{M_s} \varepsilon_{xx} \cos(\varphi_H) - 2\frac{B_2}{M_s} \varepsilon_{xz} \sin(\varphi_H) \\
\mu_0 h_{\text{me},z} &= -2\frac{B_1}{M_s} \varepsilon_{zz} \sin(\varphi_H) - 2\frac{B_2}{M_s} \varepsilon_{xz} \cos(\varphi_H)
\end{aligned}
\right..
\end{equation}

We can rotate the driving field into the local frame of reference ($x'$,$y'$,$z'$), where $x'$ is aligned with the magnetization equilibrium direction, and is aligned with the magnetic field ($\varphi_H$)

\begin{equation}\label{eq_supp:h_perp}
    \begin{pmatrix}
        h_{y'} \\ h_{z'} 
    \end{pmatrix} = 
    \begin{pmatrix}
        -\sin\varphi_H & \cos\varphi_H  & 0 \\ 0 & 0 & 1
    \end{pmatrix}
    \begin{pmatrix}
        h_{x} \\ h_{y} \\h_{z}
    \end{pmatrix} = 
    \begin{pmatrix}
        \tfrac{B_1}{2M_s}\,\varepsilon_{xx}(\omega)\sin(2\varphi_H)
        \;+\;
        \tfrac{B_2}{M_s}\,\varepsilon_{xz}(\omega)\sin^2\varphi_H \\        
        -\,\,\tfrac{B_1}{M_s}\,\varepsilon_{zz}(\omega)\sin\varphi_H
        \;+\;
        \tfrac{B_2}{M_s}\,\varepsilon_{xz}(\omega)\cos\varphi_H
        \end{pmatrix}.
\end{equation}

Now, we can replace Eq. \eqref{eq_supp:h_perp} into Eq. \eqref{eq_supp:m_prime}:

\begin{equation}\label{eq:my_lorentz}
m_{y'}(\omega) \simeq
\frac{\gamma^2 H_{\rm eff}\,h_{y'}(\omega) - i\gamma\omega\,h_{z'}(\omega)}
{D(\omega)}
\end{equation}
\begin{equation}\label{eq:mz_lorentz}
m_{z'}(\omega) \simeq
-\;\frac{\gamma^2 H_{\rm eff}\,h_{z'}(\omega) + i\gamma\omega\,h_{y'}(\omega)}
{D(\omega)}
\end{equation}
\noindent
and go back to the normal coordinate system where ($x ||$ SAW)
\begin{equation}
    \begin{pmatrix}
        m_x \\ m_y
    \end{pmatrix} = 
    \begin{pmatrix}
        \cos\varphi_H & -\sin\varphi_H \\
        \sin\varphi_H & \cos\varphi_H
    \end{pmatrix}
    \begin{pmatrix}
        m_{x'} \\ m_{y'}
    \end{pmatrix},
\end{equation}
finally reading as

\begin{equation}\label{eq_sm:m_gen}
\begin{pmatrix}
    m_x \\ m_y \\m_z
\end{pmatrix} = 
\begin{pmatrix}
    \cos\varphi_H - \sin\varphi_H m_{y'}(t) \\ 
    \sin\varphi_H + \cos\varphi_H m_{y'}(t) \\
     m_{z'}(t)
\end{pmatrix}.
\end{equation}


\subsection*{Supplementary Note 4: Acoustic orbital Hall effect (AOHE) and its back-action}

The acoustic orbital Hall effect couples the lattice dynamics to the orbital degree of freedom. Rayleigh SAWs generate a rotation of the lattice at every point in space and time, where this lattice displacement is in the $x$, $y$ and $z$-directions, as defined in Eq. \eqref{eq_supp:u}

To simplify, we can simplify and assume a fully elliptical motion ($\beta \sim 1$). Then the local rotation is defined as $\theta = \frac{1}{2} \nabla \times \mathbf{u}$ and since the displacement is in $x$ and $z$, then only the $y$-component survives:

\begin{equation} \label{eq:theta_y}
    \theta_y = \frac{1}{2}\left( \frac{\partial u_z}{\partial x} - \frac{\partial u_x}{\partial z} \right).
\end{equation}

Using Eqs.~\eqref{eq_supp:u} in \eqref{eq:theta_y}
\begin{equation}
\theta_y(x,z,t)
= -k u_0e^{kz}\cos\!\big(kx - \omega t\big).
\end{equation}

The corresponding lattice angular velocity (rotation rate) is then
\begin{equation}
\Omega_y(x,z,t) = \frac{\partial \theta_y}{\partial t}
= -k \omega u_0e^{kz}\sin\!\big(kx - \omega t\big).
\label{eq:Omega_y}
\end{equation}

When the electronic orbital is coupled to the lattice, the lattice rotation transfers angular momentum to the orbital, generating an orbital current that travels in the $z$ direction and is polarized in $y$, scaling with the local lattice angular velocity:

\begin{equation}
    j_{L,0}^{y} = A \Omega_y(x,z,t) = -A k \omega u_0e^{kz}\sin\!\big(kx - \omega t\big),
    \label{eq.sm:jl0}
\end{equation}

\noindent where A captures the efficiency with which the lattice rotation generates the orbital current, containing a phenomenological constant $\chi_{po}$ that determines the phonon-orbital coupling strength. For simplicity, in the main manuscript, we omit the decay in the $z$-direction. This orbital current can interact with an adjacent FM material and cause absorption/reflection, which can then be converted into a charge current through the IOHE. The equation that determines this interaction depends on the current from Eq.\ \ref{eq.sm:jl0} and is proportional to the torque exerted into the magnetization in the ferromagnet as suggested by \cite{TheorySpinHallMR,acousticOHE2025,Acoustic_spin_Hall_SO_saw}
\begin{equation}\label{eq:AOHE-ba}
    \mathbf{j}_{\mathrm{L}} = j_{L,0}^{y} \left[c_1\hat{y}+
c_2 \mathbf{m} \times (\mathbf{m} \times \hat{y}) +
c_3 \mathbf{m} \times \hat{y}
\right],
\end{equation}

\noindent where the coefficients $c_i$ are interface-dependent and $\alpha$ is the direction of polarization of the AOHE. Since we measure the transversal voltage ($y$-direction) we are interested in $j_{L}^x$. Expanding eq. \ref{eq:AOHE-ba} and keeping only relevant terms we obtain

\begin{equation} \label{eq:jxL}
    j_{\mathrm{L}}^x = j_{L,0}^{y} \big[ c_2 (m_y m_x)\big].
\end{equation}

Taking the magnetization components from a ME interaction from Eq. \eqref{eq_sm:m_gen}, we can substitute the product $m_y m_x \approx \sin\varphi \cos\varphi + \cos(2\varphi)\,m_{y'}$ into Eq. \ref{eq:jxL} getting
\begin{equation*}
    j_{\mathrm{L}}^x = c_2\, j_{L,0}^y\,\big[\sin\varphi_H\cos\varphi_H\big]
+ c_2\, j_{L,0}^y\,\big[\cos(2\varphi)\, m_{y'}(t)\big].
\end{equation*}

The first term, contains only the oscillation of $j_{L,0}^y$, therefore, the time-average will be zero, meanwhile the second term has $j_{L,0}^y \, m_{y'}$ which will result in a non-zero time average. We omit any $\varepsilon_{xz}$ since its out-of-phase with $j_{L,0}$ and the time average will give zero ($\left<\sin(kx-wt)\cos(kx-wt)\right>=0$). Therefore, $j_{\mathrm{L}}^x$ is described by

\begin{equation}
     j_{\mathrm{L}}^x  = -j_{L,0}(t)\frac{\, c_2\,}{D(\omega)}\left[ \frac{\gamma^2 H_{\rm eff}B_1\varepsilon_{xx}(t)}{2M_s} \sin(4\varphi) + \frac{i\gamma B_1 \beta\, \varepsilon_{zz}(t)}{M_s} \cos(2\varphi_H)\sin(\varphi_H)\right],
\end{equation}

\noindent where the terms $\varepsilon_{xx}$ and $\varepsilon_{zz}$ are in phase with $j_{L,0}$ and will give a non-zero time average ($\left<\sin^2(kx-wt)\right>=1/2$). We can further simplify taking $\beta = 0.5$. Moreover, we are only interested in the real part {$\Re(j_{l,0}\, m_{y'}^*)$}. We can substitute Eq. \eqref{eq.sm:jl0}-\eqref{eq_sm:strain} and take $1/D^*(w) = \frac{(\omega_r^2-\omega^2)+i\omega\Gamma}{(\omega_r^2-\omega^2)^2-(\omega\Gamma)^2}$, 

\begin{equation}
    \langle j_{\mathrm{L}}^x \rangle_t = -\frac{c_2Ak^2u_0^2e^{2kz}\gamma B_1}{4M_s[(\omega_r^2-\omega^2)^2-(\omega\Gamma)^2]} \left[\gamma H_{\rm eff}(\omega_r^2-\omega^2)  \sin(4\varphi_H) + \omega\Gamma  \cos(2\varphi_H)\sin(\varphi_H)\right].
\end{equation}

The first term is dispersive/anti-symmetric Lorentzian as also found in Ref. \cite{acousticOHE2025}, and the second one is an absorptive/symmetric Lorentzian. Then, through the IOHE, the orbital current is converted into a measurable charge current in the $y$-direction and the voltage will be


\begin{equation}
    V_{xy}^{\rm AOHE} \propto  \theta_{OH}\rho L_y\langle j_{\mathrm{L}}^x \rangle_t,
\end{equation}

\noindent where $\theta_{OH}$ is the orbital Hall angle, $\rho$ is the sample's resistivity and $L_y$ is the sample's len. The second term of $\langle j_{\mathrm{L}}^x \rangle_t$ results in an asymmetry between the two first and the two last peaks. In our main manuscript and on Supplementary Note 6, we show different curves that don't show this asymmetry and our voltage is just  proportional to $\sin(4\varphi_H)$. Note that the AOHE voltage depends on the local magnetization state, which is enhanced at the resonant condition, thus also enhancing the AOHE signal. 


\subsubsection*{Standing wave configuration}
The displacement equation are modified by summing a forward and backward propagating wave: 
\begin{align}
u_x(x,t) &= u_x^+(x,t) + u_x^-(x,t)
= 2 u_0  e^{k z} \cos(kx)\cos(\omega t), \\
u_z(x,t) &= u_z^+(x,t) + u_z^-(x,t)
= -2 \beta u_0 e^{k z}  \sin(kx)\sin(\omega t),
\end{align}
where $u_0$ is the amplitude of the forward wave and $\alpha$ is the ellipticity parameter.

\begin{align}
\varepsilon_{xx} &= \frac{\partial u_x}{\partial x} 
= -2 k u_0 e^{k z} \sin(kx) \cos(\omega t), \\[2mm]
\varepsilon_{zz} &= \frac{\partial u_z}{\partial z} 
= -2 \beta k u_0 e^{k z} \sin(kx) \sin(\omega t), \\[1mm]
\varepsilon_{xz} &= \frac{1}{2} \left( \frac{\partial u_x}{\partial z} + \frac{\partial u_z}{\partial x} \right) 
=  k u_0 e^{k z} \cos(kx) \left[ \cos(\omega t) - \beta \sin(\omega t) \right].
\end{align}

Following the same steps as before, the AOHE current density is

\begin{equation}
    \langle j_{\mathrm{L}}^x \rangle_t = -\frac{c_2Ak^2u_0^2e^{2kz}B_1}{M_s} \left[\frac{\gamma^2 H_{\rm eff}(\omega_r^2-\omega^2)  }{4 [(\omega_r^2-\omega^2)^2-(\omega\Gamma)^2]} \beta \sin(kx) \sin(4\varphi_H)\right],
\end{equation}

Note that this expression has an additional spatial dependence ($\sin(kx)$), which must be integrated over the length of the sample ($L_x$) to obtain the total current:

\begin{align}
I &= \int_0^{L_x} \langle j_{\mathrm{L}}^x(x) \rangle_t \, dx
\end{align}

Depending on the wavelength ($\lambda = 2\pi/k$) of the standing wave, if ($L_x$) is not an integer multiple of ($\lambda$), the integral is non-zero, resulting in a finite detected AOHE current.


\subsection*{Supplementary Note 5: Acoustic pumping}

The acoustic pumping comes from the magnetization dynamics that is driven by the magnetoelastic effect that generates a magnetoacoustic wave in the FM. The pumping equation can be written as follows
\begin{equation}\label{eq:inst-js}
  \mathbf{j}_{AP}(t) \;=\; \frac{\hbar}{4\pi}\,g_{\uparrow\downarrow}\,\mathbf{m}(t)\times\dot{\mathbf{m}}(t),
\end{equation}

\noindent where $\hbar$ is Plancks constant, $g_{\uparrow\downarrow}$ is the spin or orbital transparency at the interface, $\textbf{m}$ is the normalized magnetization and $\dot{\textbf{m}}$ is the magnetization dynamics governed by the Landau-Lifshitz-Gilbert equation.
We define a local frame in which $x' = m_0$, where $m_0$ is the equilibrium position, $y'$ is in-plane transverse to $m_0$, and $z'=z$ (out-of-plane). We also assume small amplitude variations around the equilibrium position and we can define the cross product leading (second) order in the small amplitude, the time-average of \(\mathbf{m}\times\dot{\mathbf{m}}\) is
\[
\langle \mathbf{m}\times\dot{\mathbf{m}} \rangle
\approx \langle \delta\mathbf{m}\times\dot{\delta\mathbf{m}} \rangle.
\]

\noindent as the variations occur only in the transverse directions, we can write the pumping as 

\begin{equation}\label{eq:js-mag}
  \langle j_{AP} \rangle \;=\; \frac{\hbar}{4\pi}\,g_{\uparrow\downarrow}\,\omega\,|\mathbf{m}_\perp|^2.
\end{equation}

Using \eqref{eq:js-mag} and that $\mathbf{m}_\perp = m_{y'}+m_{z'}$, the time-averaged pumped spin current magnitude is
\begin{equation}\label{eq:js_full}
  \langle j_{AP}\rangle \;=\; \frac{\hbar}{4\pi}\,g_{\uparrow\downarrow}\,\omega|m_{y'}+m_{z'}|^2 .
\end{equation}

Substitute the \(m_{y'}\) [Eq.~\eqref{eq:my_lorentz}] and \(m_{z'}\) [Eq.~\eqref{eq:mz_lorentz}]. Expanding the squared magnitude and considering that the driving fields are out-of-phase, the equation is finally

\begin{equation}\label{eq:js_xx}
  j_{AP} = \frac{\hbar}{4\pi}g_{\uparrow\downarrow}\omega\;\frac{(\gamma^2 H_{\rm eff})^2 + (\gamma \omega)^2}{|D(\omega)|^2} (|h_{y'}|^2+|h_{z'}|^2).
\end{equation}

Keeping only the non-zero time-average terms ($|\varepsilon_{xx}|^2$, $|\varepsilon_{zz}|^2$, $|\varepsilon_{xz}|^2$):

\begin{align}
|h_{y'}|^2 &=
\left(\frac{B_1}{2M_s}\right)^2 |\varepsilon_{xx}|^2 \sin^2 2\varphi_H
\;+\;
\left(\frac{B_2}{M_s}\right)^2 |\varepsilon_{xz}|^2 \sin^4\varphi_H,
\\[6pt]
|h_{z'}|^2 &=
\left(\frac{B_1}{M_s}\right)^2 |\varepsilon_{zz}|^2 \sin^2\varphi_H
\;+\;
\left(\frac{B_2}{M_s}\right)^2 |\varepsilon_{xz}|^2 \cos^2\varphi_H.
\end{align}

Finally, plugging all values, we get

\begin{align}
\langle j_{AP} \rangle &= 
\frac{\hbar}{4 \pi} g_{\uparrow\downarrow} \,\omega \,
\frac{ (\gamma^2 H_{\rm eff})^2 + (\gamma \omega)^2 }{|D(\omega)|^2} \; \times \nonumber\\
&\quad \Bigg[
\frac{B_1^2}{4 M_s^2} (k u_0)^2 \sin^2 2\varphi_H
+ \frac{B_2^2}{4 M_s^2} (k u_0)^2 (1-\beta)^2 \sin^4 \varphi_H \nonumber\\
&\qquad + \frac{B_1^2}{M_s^2} (\beta k u_0)^2 \sin^2 \varphi_H
+ \frac{B_2^2}{4 M_s^2} (k u_0)^2 (1-\beta)^2 \cos^2 \varphi_H
\Bigg].
\end{align}


Then its conversion to a charge current will be through ISHE or IOHE as

\begin{equation}
  \mathbf{E}_{\rm ISHE/IOHE} \;=\; \theta_{\rm SH/OH}\,\rho\,\langle \mathbf{j}_{AP}\rangle \times \hat\sigma.
\end{equation}

which projected to the magnetization equilibrium direction ($V_{xy}$) is:
\[
  E_y^{\rm HE} \;=\; \theta_{\rm SH/OH}\,\rho\,\langle j_{AP}\rangle\;(\hat z\times\hat\sigma)\cdot\hat y
  \;=\; \theta_{\rm SH/OH}\,\rho\,\langle j_s\rangle\cos\theta,
\]

The final angularly-resolved voltages in the transversal direction are:
\begin{align}
 V_{xy}^{AP} &= \theta_{SH/OH} \rho L_y
\frac{\hbar}{4 \pi} g_{\uparrow\downarrow} \,\omega \,
\frac{ (\gamma^2 H_{\rm eff})^2 + (\gamma \omega)^2 }{|D(\omega)|^2} \; \times \nonumber \\
&\quad \Bigg[
\frac{B_1^2}{4 M_s^2} (k u_0)^2 \sin^2 2\varphi_H
+ \frac{B_2^2}{4 M_s^2} (k u_0)^2 (1-\beta)^2 \sin^4 \varphi_H \nonumber\\
&\qquad + \frac{B_1^2}{M_s^2} (\beta k u_0)^2 \sin^2 \varphi_H
+ \frac{B_2^2}{4 M_s^2} (k u_0)^2 (1-\beta)^2 \cos^2 \varphi_H
\Bigg] \cos(\varphi_H).
\end{align}


A typical value of ellipticity in LiNbO$_3$ is $\beta\approx 0.5$; therefore, the voltage will be dominated by

\begin{align}
 V_{xy}^{AP} &= \theta_{SH/OH} \rho L_y
\frac{\hbar}{4 \pi} g_{\uparrow\downarrow} \,\omega \,
\frac{ (\gamma^2 H_{\rm eff})^2 + (\gamma \omega)^2 }{|D(\omega)|^2} \frac{(B_1\,k u_0)^2 }{4 M_s^2}  \Bigg[\sin^2 2\varphi_H+ \sin^2 \varphi_H\Bigg] \cos(\varphi_H).
\label{eq:Vy_xx}
\end{align}



\subsection*{Supplementary Note 6: Frequency and power dependence of AP and AOHE in Ni$|$Cr}
\begin{figure}[htb!]
    \centering
    \includegraphics[width=0.6\columnwidth]{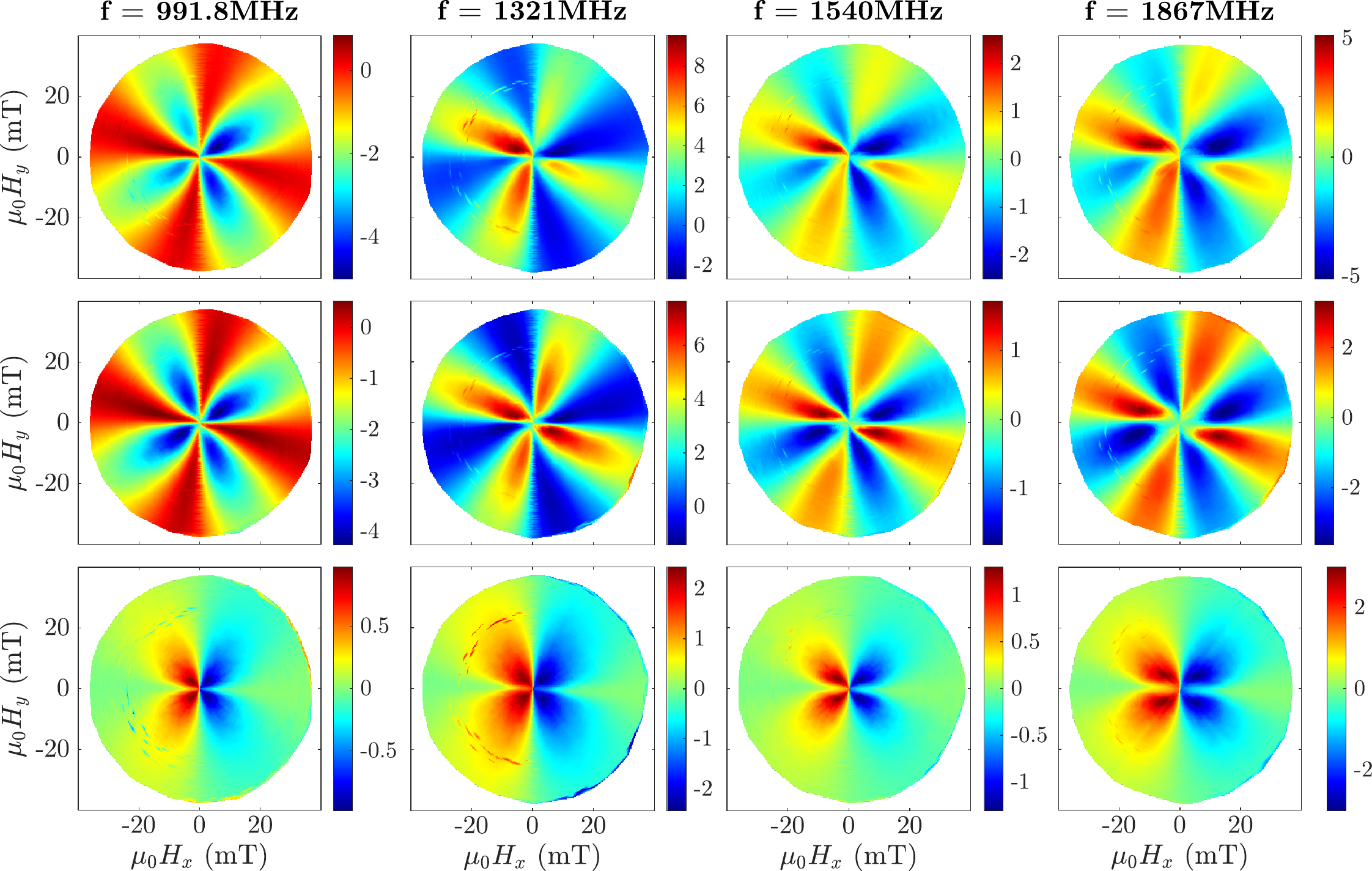}
    \caption{Voltage maps as a function of magnetic-field angle and magnitude at four different frequencies. Each column corresponds to a frequency, increasing from 991.8 MHz to 1867 MHz (left to right). The rows show the total signal (top), and its even (middle) and odd (bottom) components. Each panel includes its own colorbar indicating the measured voltage amplitude.
    }
    \label{fig_supp:vsfreq}
\end{figure}
Our experiment involved a systematic frequency-dependent study using efficient IDTs operating from 111 MHz up to 2 GHz. Figure~\ref{fig_supp:vsfreq} displays voltage maps for four representative frequencies (each column), showing the total signal (top), together with its even (middle) and odd (bottom) components. As the frequency increases, the magnetic resonance shifts to higher fields. This appears in the odd component as an outward displacement of the characteristic lobes. We note also that in the even component, there is vanishing signal at low fields as the frequency increases, which can be attributed to the system entering resonance and developing a phase shift between the magnetization oscillation and the SAW-generated orbital current in the light metal.

In the total signal, we observe that increasing frequency progressively makes the angular pattern resemble that of SAW-driven pumping, while at lower frequencies the AOHE contribution is more prominent. This trend is analyzed in more detail in Fig.~\ref{vsfreq_vspower}.

\begin{figure}[htb!]
    \centering
    \includegraphics[width=0.8\columnwidth]{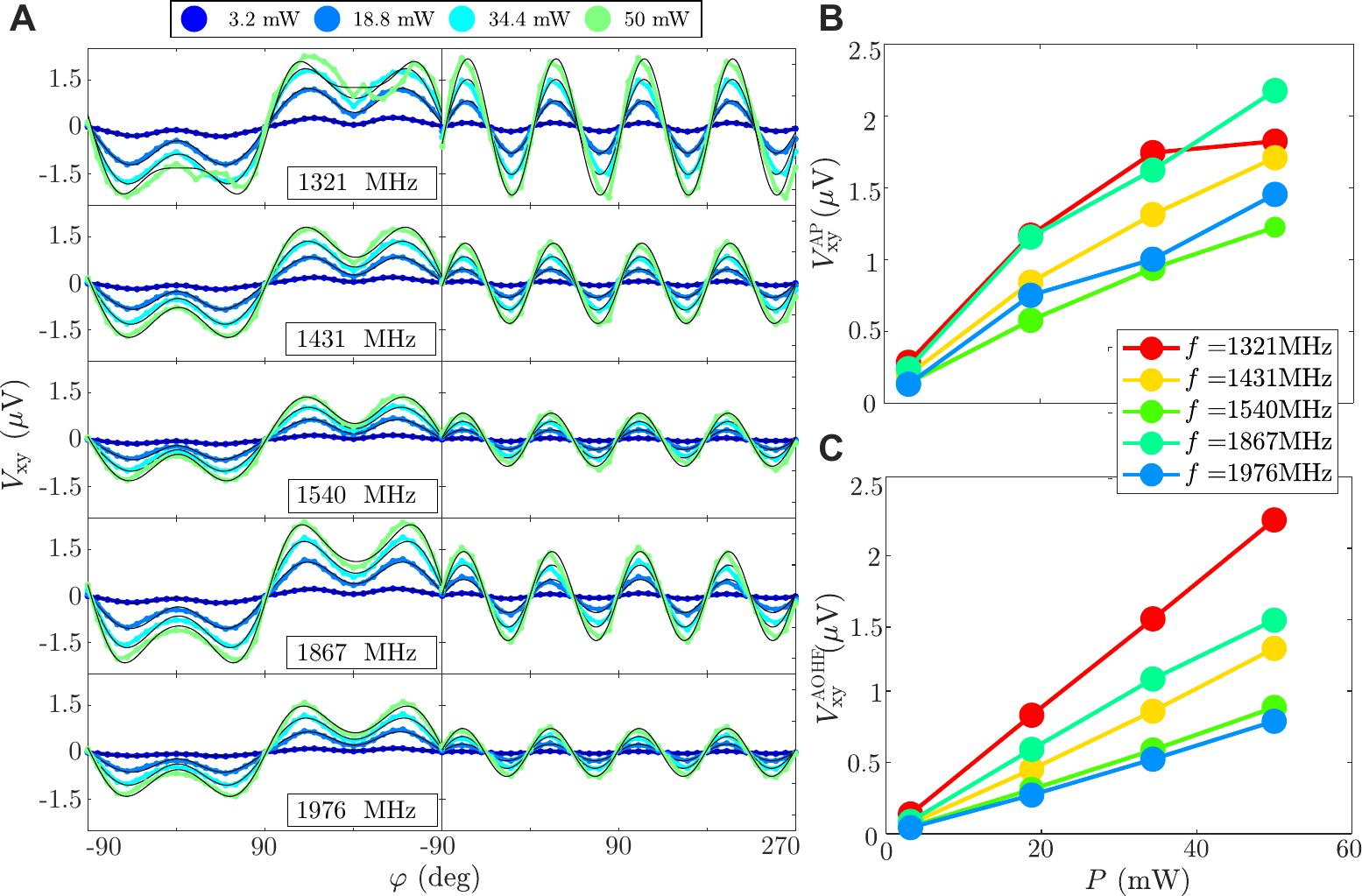}
    \caption{
    (\textbf{A}) Left column shows the AP voltage taken at 3.5 mT at different frequencies, and the color of each curve corresponds to the input power of the IDT. The right column displays the AOHE voltage taken at 35 mT for different frequencies and input power. 
    (\textbf{B}) Summary of the AP amplitude versus input power of the IDT. Each color represents a different frequency. \textbf{(C)} Summary of the AOHE amplitude versus the input power of the IDT. Each color represents a different frequency, and they all follow a linear trend.
    }
    \label{vsfreq_vspower}
    
\end{figure}

Figure\ \ref{vsfreq_vspower}a shows the power dependence at diferent IDT frequencies (each row), from 1321 to 1976 MHz. The field-odd component associated to the acoustic pumping, $V_{\rm AP}$ in the left panels and the field-even component associated with AOHE, $V_{\rm AOHE}$, in the right-hand-side panels. The magnetic field for the AP is taken at the resonance (which varies at each around 2-4 mT), while for AOHE its taken at 35 mT. As expected, for all frequencies and in both AP and AOHE, we see an upward trend as we increase the power. Frequencies cannot be compared directly since their efficiency depends on the IDT and as shown in Fig.\ \ref{setup_dim} some frequencies transmit better than others. We observe a linear increase with power at all frequencies in the AOHE (Fig.\ \ref{vsfreq_vspower}c), whereas AP (Fig.\ \ref{vsfreq_vspower}b) shows signs of saturation, especially for the most efficient frequencies (1321 MHz).


We note first that the assumption in determining the AP voltage in Eq.\ \ref{eq:Vy_xx} is made for linearized LLG equation, which works well for small precession angles of magnetization. We know SAW could generate large amplitude magnetization precession and we indeed observe a  light change in the AP voltage shape for the angular dependence as a function of power. Additionally, we may add extra terms in our magnetization dynamics terms ($m_{y'}$) arising from the torque exerted by the AOHE generated in the LM. The orbital current is polarized in the $y$-direction, and may induce field-like and damping-like torques on the FM's magnetization, coupling with the magnetoelastic effect, and giving extra terms and account for this change in symmetry.



\subsection*{Supplementary Note 7: Second Harmonic }

Current-induced orbital torques in our FM$|$LM bilayers were also analyzed by second Harmonic transport experiments in Hall bar structures \cite{Manchon2019}. A 10 Hz ac current $I$ is applied to the device and the first and second harmonic Hall voltage responses $V_{xy}^{1\omega}$ and $V_{xy}^{2\omega}$ are extracted by demodulating the signal at the excitation frequency. Figures\ \ref{fig_supp:sot}a and \ \ref{fig_supp:sot}b show the first and second harmonic Hall resistance $R_{xy}^{1\omega}=V_{xy}^{1\omega}/I$ and $R_{xy}^{2\omega}=V_{xy}^{2\omega}/I$ measured in a Ni(15)$|$Ti(8) device capped with 3nm-thick-Al layer.

The first harmonic resistance response captures the anomalous Hall effect of the Ni following a $\sin2\phi$ angular dependence as expected (Figure\ \ref{fig_supp:sot}a), where $\phi$ defines the in-plane angle between the magnetic field and the current line. The second harmonic response is given by \cite{Manchon2019}

\begin{equation}
    R_{xy}^{2\omega} = (-\frac{1}{2}R_{xy,DL}^{2\omega} +R_{xy,SSE}^{2\omega}-R_{xy,FL}^{2\omega})\cos\phi +2R_{xy,FL}^{2\omega}\cos^3\phi,
    \label{eq:2wSOT}
\end{equation}

\noindent
where $R_{xy,DL}^{2\omega}$, $R_{xy,FL}^{2\omega}$ and $R_{xy,SSE}^{2\omega}$ represent the amplitude of the damping-like (DL) torque, field-like (FL) torque, and spin Seebeck effect (SSE) contributions. Both $R_{xy,DL}^{2\omega}$ and $R_{xy,FL}^{2\omega}$ decay with the effective magnetic field $B_{eff}$ as $1/B_{eff}$, whereas $R_{xy,SSE}^{2\omega}$ can be considered as a constant as long as the magnetization of the FM is saturated. The effect of the spin/orbital currents generated from charge flow in the LM is captured in $R_{xy,DL}^{2\omega}$, while $R_{xy,FL}^{2\omega}$ originates from the combination of interface effects and the current-induced Oerstead field. According to its origin, $R_{xy,DL}^{2\omega}=R_{xy,AHE}^{1\omega}\frac{B_{DL}}{B_{eff}}$, where $R_{xy,AHE}$ is the anomalous Hall effect (AHE) of the FM and $B_{DL}$ the DL magnetic field generated by the current. 

The different contributions to $R_{xy}^{2\omega}(\phi)$ can be extracted by performing angular dependent measurements at different magnetic field strengths such as the ones shown in Fig.\ \ref{fig_supp:sot}b, and fitting them to Eq.\ \ref{eq:2wSOT}. In particular, $B_{DL}$ is extracted by subtracting the FL contribution to the $\cos\phi$ term in Eq. \ \ref{eq:2wSOT}, and analyze its magnetic field dependence as shown in Fig. \ \ref{fig_supp:sot}c. Accordingly, the slope of the linear fit to the data yields the estimate of $B_{DL}$. 

The same current-induced torque analysis was performed for all the bilayers investigated in this work, and confirmed its origin by verifying the expected linear dependence of $B_{DL}$ with $I$. The results obtained for all Ni-based samples are summarized in Fig.\ \ref{fig_supp:sot}d. For the case of Co$|$LM samples, the DL-torque contribution to $R_{xy}^{2\omega}$ was too weak to quantify the amplitude and sign of $B_{DL}$. Therefore, we conclude that the DL torque in the Co-based samples was negligible compared to Ni samples. The difference between the Ni and Co samples, together with the analysis of the torques estimated by STT-FMR corroborate that the torques originate from current-indcued orbital moments. Besides, the qualitative agreement between SOT, STT and AP signals confirms their underlying common physical origin, i.e., OHE or IHOE, whereas the SAW-driven AOHE is of a different nature.

\begin{figure}[htb!]
    \centering
    \includegraphics[width=0.8\columnwidth]{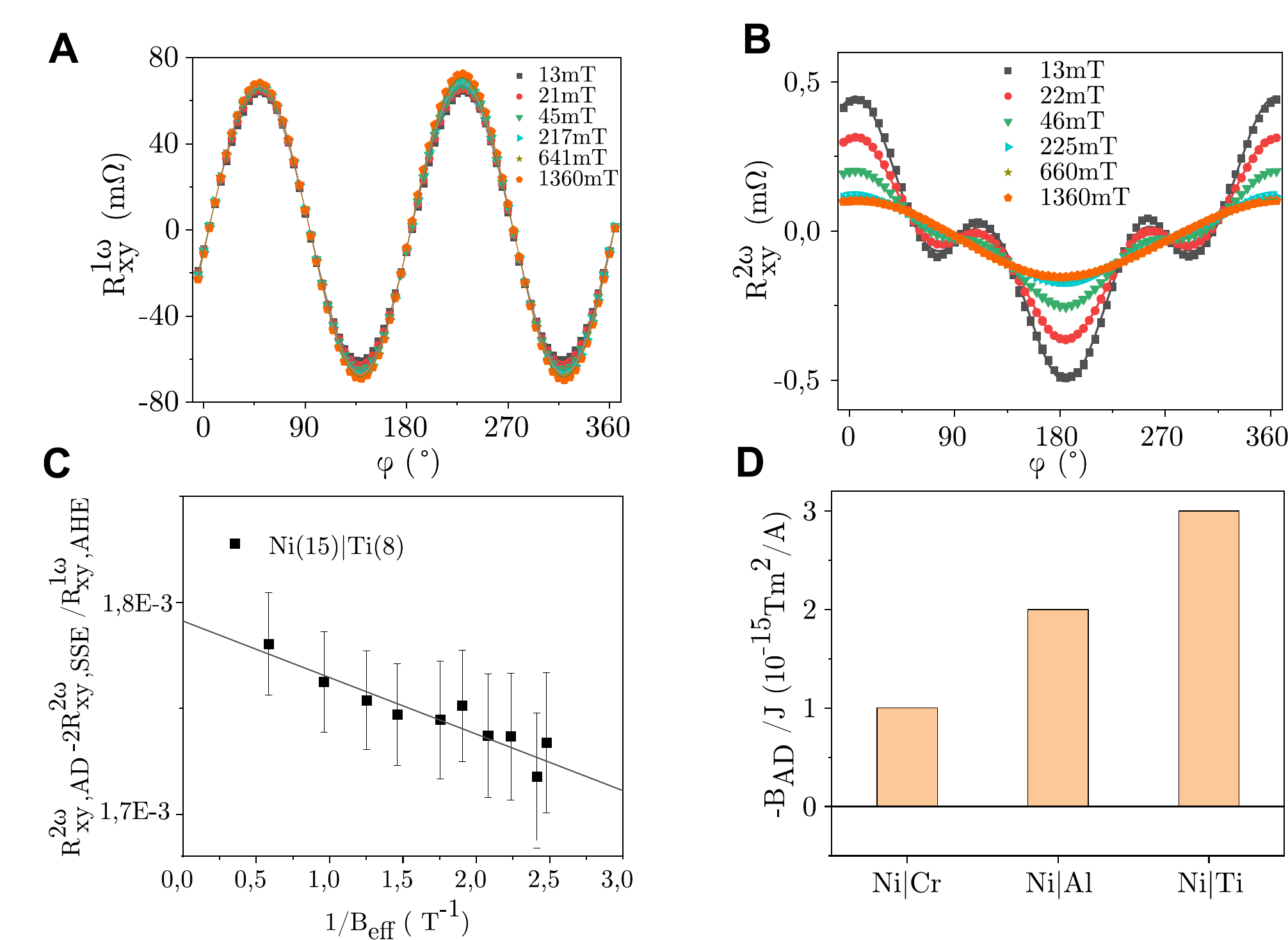}
    \caption{
    (\textbf{A}) $R_{xy,DL}^{1\omega}$ and
    (\textbf{B}) $R_{xy,DL}^{2\omega}$ as a function of in-plane angle $\phi$ obtained in Ni(15)$|$Ti(8) at different magnetic field strengths.
    (\textbf{C}) Analysis of the DL torque in Ni(15)$|$Ti(8) extracted from the fit of the data in (B) to Eq.\ \ref{eq:2wSOT}. $R_{xy,AHE}$ is evaluated from AHE measurements. $B_{eff}$ is the sum of the external applied field and the anisotropy field of the FM layer, which is evaluated from $R_{xy}(H)$ measurements. 
    (\textbf{D}) Summary of the $B_{DL}$ values extracted from current-induced torque analysis in all Ni-based samples. The data is normalized by the current density $J$ applied to the LM. 
    }
    \label{fig_supp:sot}
\end{figure}


\subsection*{Supplementary Note 8: STT-FMR and OTT-FMR}
Spin-transfer torque (STT) and orbital-transfer torque (OTT) are the reciprocal processes of spin and orbital pumping. Here, an electrical current is injected into the non-magnetic (NM) layer, where it is converted into a spin or orbital current through the spin Hall or orbital Hall effect. When this spin/orbital current enters the ferromagnet (FM), it exerts a torque on the magnetization and drives its precession.

In addition to this torque, the injected RF current also generates an accompanying electric and magnetic field—commonly referred to as the Oersted field. This RF excitation behaves like a coplanar waveguide and can independently drive ferromagnetic resonance (FMR) in the FM layer. In STT-FMR measurements, these two contributions are separated by fitting the measured voltage to a combination of symmetric and antisymmetric Lorentzian components. The antisymmetric part arises from the Oersted field and other field-like torques, whereas the symmetric part originates from the spin/orbital transfer torque, also known as the damping-like torque.
In our experiment, we performed STT-FMR between 5 and 10 GHz for all samples. From these measurements, we extract the torque efficiency, which reflects the strength of the spin/orbital torque. The symmetric and antisymmetric contributions to the total voltage are given by:

\begin{align}
V_{\text{sym}} &= \frac{I_{\text{RF}} \Delta R}{2} \mu_0 H_{\text{DL}} \sin 2\theta \cos \theta \, 
\frac{\mu_0 H_{\text{res}} + \mu_0 M_{\text{eff}}}{W (2 \mu_0 H_{\text{res}} + \mu_0 M_{\text{eff}})} 
\sqrt{\frac{\mu_0 H_{\text{res}}}{\mu_0 M_{\text{eff}} + \mu_0 H_{\text{res}}}}, \\[1em]
V_{\text{antisym}} &= \frac{I_{\text{RF}} \Delta R}{2} \left( \mu_0 H_{\text{Oe}} + \mu_0 H_{\text{FL}} \right) 
\sin 2\theta \cos \theta \, 
\frac{\mu_0 H_{\text{res}} + \mu_0 M_{\text{eff}}}{W (2 \mu_0 H_{\text{res}} + \mu_0 M_{\text{eff}})}.
\end{align}

\noindent
Here, $I_{RF}$ is the injected current which we use an rf power of $P=8$ dBm, $\Delta R$ is the magnetoresistance, $\mu_0 H_{DL}$ is the effective damping-like field, $\mu_0H_{\rm res}$ is the resonant field at the studied frequency (6 GHz) and $\mu_0 M_{\rm eff}$ is the effective magnetization of each sample. A commonly used first estimate of the torque efficiency is obtained from the ratio of the symmetric and antisymmetric components of the ST-FMR lineshape. This ratio provides an overall measure of how much damping-like torque is present relative to the combined field-like and Oersted contributions. In this simplified approach, the efficiency can be written as

\begin{equation}
    \frac{V_{\rm sym}}{V_{\rm asym}} = \frac{\mu_0 H_{\rm DL}}{\mu_0H_{\rm Oe}+\mu_0H_{\rm FL}} \sqrt{\frac{\mu_0 M_{\rm eff}}{\mu_0 M_{\rm eff}+\mu_0 H_{\rm res}}}
\end{equation}

\begin{figure}[ht]
    \centering
    \includegraphics[width=0.7\columnwidth]{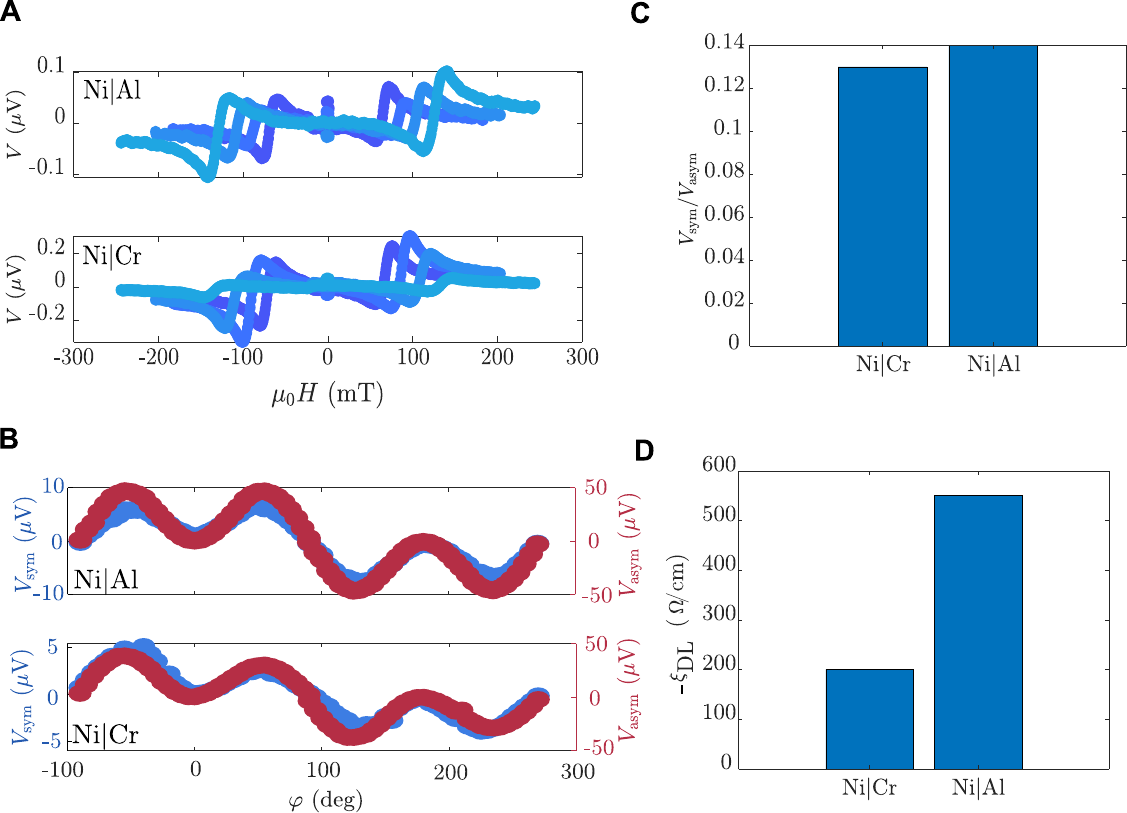}
    \caption{
    (\textbf{A}) Voltage curves at different excitation frequencies between (5-8 GHz) for both samples, from top to bottom: Ni$|$Al, Ni$|$Cr. 
    (\textbf{B}) voltage amplitude of the symmetric (blue) and anti-symmetric (red) Lorentzian as a function of the angle between the input current ($I_{ac}$) and the magnetic field. 
    (\textbf{C}) Quotient between the $V_{\rm sym}$ and  $V_{\rm asym}$ that is related to the efficiency of the damping-like torque for all samples.
    (\textbf{D}) Efficiency of the damping-like effective field, from only the symmetric voltage amplitude($V_{\rm sym}$).    
    }
    \label{STTFMR}
\end{figure}

Figure~\ref{STTFMR}a displays the STT-FMR curves between 5 and 8~GHz for all samples. Capping Ni with Cr or Al does not significantly modify the line shape. To compare amplitude and symmetry across samples, we performed angle-dependent STT-FMR at 6~GHz. Figure~\ref{STTFMR}b shows $V_{\mathrm{sym}}$ (left axis) and $V_{\mathrm{asym}}$ (right axis) as a function of the angle between the RF current and the static magnetic field. All samples follow the expected $\sin(2\theta)\cos\theta$ dependence and show a dominant antisymmetric contribution. The ratio $V_{\mathrm{sym}}/V_{\mathrm{asym}}$, plotted in Fig.~\ref{STTFMR}c, again shows no significant difference between Cr and Al samples.

This lineshape ratio does not isolate the damping-like torque completely, but it provides a useful relative indication of how the torque changes between samples. We can go one step further and follow Ref. \cite{Hayashi2023_st_fmr_niTi} to extract the pure damping-like effective field $H_{DL}$ from the symmetric component alone with:

\begin{equation}
    \mu_0 H_{\rm  DL} = \frac{2SW}{I_{RF}\Delta R}\frac{2\mu_0H_{\rm res}+\mu_0M_{\rm eff}}{\mu_0H_{\rm res}+\mu_0M_{\rm eff}} \sqrt{1+\frac{\mu_0M_{\rm eff}}{\mu_0H_{\rm res}}},
\end{equation}

where $S$ is the result of the fit on the angular-symmetric contribution to the STT-FMR experiment (see Fig. \ref{STTFMR}), and $W$ is the full width at half maximum (FWHM). Taking all the previous parameters, we can find the efficiency of the damping-like field for each sample with the following equation:


\begin{equation}
    \xi_{DL}^E = -\left( \frac{2e}{\hbar} \right)\frac{\mu_0M_st_{FM}H_{DL}}{E},
\end{equation}

\noindent
where $t_{FM} = 10$ nm is the thickness of the ferromagnetic film, $H_{DL}$ is the damping-like field in Oersted, and $E$ is the electric field due to the injected current and is calculated as $E=I_{RF}R_0/d$, where $R_0$ is the resistance of the sample and $d$ is the distance between the golden pads, in this case around 400 $\mu$m. The results are displayed in Table \ref{tab:samples}.

\begin{table}[h!]
\caption{Magnetic and electrical parameters of the different samples used to obtained the STT-FMR efficiencies.}
\centering
\rowcolors{2}{gray!10}{white}
\begin{tabular}{l 
    S[table-format=1.2] 
    S[table-format=3.0] 
    S[table-format=1.3] 
    S[table-format=1.3] 
    S[table-format=1.3]}
\toprule
\textbf{Sample} & {$\mu_0 M_{\text{eff}}$ (T)} & {$\mu_0H_{\rm res}$ (mT)}& {$\Delta R$ (\si{\ohm})} & $V_{\rm sym}/V_{\rm asym}$ &  {$\xi_{DL}$ ($\Omega/$cm)}\\
\midrule
Ni$|$Cr & 0.63 & 85.7 & 0.100 & 0.14 & -550.3 \\
Ni$|$Al & 0.60 & 89.3 & 0.047 &  0.13 & -200.3 \\
\bottomrule
\end{tabular}

\label{tab:samples}
\end{table}




\clearpage 



\end{document}